\documentclass[11pt,reqno,a4paper]{amsart}
\usepackage[margin=2.5cm]{geometry}
\RequirePackage{amsthm,amsmath,amsfonts}
\RequirePackage[numbers]{natbib}
\RequirePackage[colorlinks,citecolor=blue,urlcolor=blue]{hyperref}
\usepackage{booktabs,longtable}
\usepackage{graphicx}
\usepackage{xcolor}
\usepackage{epstopdf}
\usepackage{physics}
\usepackage{bm}
\usepackage{hyperref}

\newtheorem{theorem}{Theorem}

\usepackage{tikz}

\newcommand{\Din}{D^{\rm in}} 
\newcommand{\Dout}{D^{\rm out}} 
\newcommand{\bftheta}{\bm{\theta}} 
\newcommand{\deltain}{\delta_{\rm in}} 
\newcommand{\deltaout}{\delta_{\rm out}} 
\newcommand{\hyb}{\mathcal{H}} 
\newcommand{\field}{\mathcal{F}} 
\newcommand{\Prob}{\mathbb{P}} 
\newcommand{\given}{\, | \,} 
\newcommand{\Expe}{\mathbb{E}} 
\newcommand{\tdeltain}{\tilde{\delta}_{\rm in}} 
\newcommand{\almostsure}{\, \overset{a.s.}{\longrightarrow} \,} 
\newcommand{\indicator}{\mathbb{I}}

\newcommand{\convas}{\, \overset{a.s.}{\longrightarrow} \,}

\newcommand{\bE}{\bm{E}}
\newcommand{\bfpsi}{\bm{\psi}}
\newcommand{\bfepsilon}{\bm{\epsilon}}

\graphicspath{{./}{../figures/}}

\begin{document}
\begin{center}
	{\Large \bf  
	Directed Hybrid Random Networks 
	Mixing Preferential Attachment with Uniform 
	Attachment Mechanisms}
	
	\bigskip
	{Tiandong Wang\footnote{\scriptsize Department of 
	Statistics, Texas 
	A\&M University, College 
		Station, TX 77843, U.S.A. Email: 
		\href{mailto:twang@stat.tamu.edu}{twang@stat.tamu.edu}}
	\qquad and \qquad Panpan Zhang\footnote{\scriptsize Department 
	of 
	Biostatistics, Epidemiology and Informatics, University of 
	Pennsylvania, Philadelphia, PA 19104, U.S.A. Email: 
		\href{mailto:panpan.zhang@pennmedicine.upenn.edu}
		{panpan.zhang@pennmedicine.upenn.edu}}	
	}
		
	\bigskip
	
	\today
\end{center}

\bigskip\noindent
{\bf Abstract.}
Motivated by the complexity of network data, we propose a 
directed hybrid random network that mixes preferential 
attachment (PA) rules with uniform attachment (UA) rules. 
When a new edge is created, with probability $p\in [0,1]$, 
it follows the PA rule. Otherwise, this new edge is added 
between two uniformly chosen nodes. Such mixture makes the 
in- and out-degrees of a fixed node grow at a slower rate, 
compared to the pure PA case, thus leading to lighter 
distributional tails. 
Under this hybrid model, however, existing inference methods (cf. 
\cite{Wan2017})
become inapplicable. Instead, we develop alternative approaches 
which 
are then applied to both synthetic and 
real datasets. We see that with extra flexibility given by 
the parameter $p$, the hybrid random network provides a 
better fit to real-world scenarios, where lighter tails from 
in- and out-degrees are observed.


Secondary: 05C07


\section{Introduction}
\label{Sec:Intro}

The {\em preferential attachment} (PA) 
mechanism~\citep{Barabasi1999} has been widely used to model 
interactions or communications among the entities in a network-based 
system, especially evolving networks. A precursory study of PA 
networks was conducted by~\citet{deSollaPrice1965} to model the 
growth of citation networks, where the research outcome coincides
with a sociological theory called the {\em Matthew 
	Effect}~\citep{Merton1968}, inducing a well known economic 
manifestation---``The rich get richer; the poor get poorer''. 

One of the most appealing properties of the PA network is {\em 
	scale-free} (i.e., the node degree distribution follows a {\em 
	power 
	law}), rendering that the PA rule has become an attractive 
choice 
for real network modeling, such as the World Wide 
Web~\citep{Henzinger2004} and collaboration 
networks~\cite{Newman2001}. We refer the readers 
to~\citet{Durrett2006, vanderhofstad2017} for some text-style 
elaborations of the elementary descriptions and probabilistic 
properties of PA networks. 
Recent studies have extended classical PA 
networks to directed counterparts, where some asymptotic 
theories~\citep{Wang2015, wang:resnick:2016, wang:resnick:2019} and 
the {\em maximum likelihood 
	estimators} (MLEs) of the parameters~\citep{Wan2017} have been 
developed. Other recent works on the mathematical treatments of PA 
networks and their variants include~\citet{Gao2017, Alves2019, 
	Mahmoud2019, Wang2020, Zhang2020}.

However, classical (either directed or undirected) PA networks do 
not always 
fit the real network data well, nor are they able to precisely 
capture some key 
attributes of the networks. Alternatively,~\citet{Atalay2011} 
proposed a model mixing PA and {\em uniform attachment} (UA) to 
investigate the buyer-supplier network in the United 
States, showing that the proposed model has outperformed the pure PA 
model. In this paper, we consider a class of directed {\em hybrid 
	random networks} (HRNs) presenting PA and UA mechanisms 
simultaneously, 
governed by a tuning parameter $p\in [0,1]$. The presence of UA 
in the proposed model effectively leverages the heavy tail produced 
by the PA mechanism, rendering that the model tentatively better 
fits the real networks whose degree distributions are less heavier.

Note that by \cite{dereich2009}, an undirected sublinear PA model
with attachment probability proportional to $\text{Degree}_v^\alpha$,
$\alpha\in (0,1)$, produces a
degree distribution with stretched exponential tails,
\[
\exp(-k^{\alpha}), \qquad \alpha\in (0,1), \qquad\text{fro k large},
\]
which is lighter than the power-law tail. With a stretched 
exponential tails,
widely adopted methods, e.g. \cite{clauset:shalizi:newman:2009}, 
may provide inaccurate tail estimates,
since the underlying distributional tail deviates from power laws;
see \cite{drees:janssen:resnick:wang:2020} for a detailed 
discussion. 
Therefore, in this paper, we restrict ourselves
to the mixture of PA and UA schemes to obtain a lighter tail of the 
degree distribution
and retain the asymptotic power-law behavior simultaneously.

In the literature, there is a limited amount of work on the random 
structures that integrate PA and UA during the 
evolution.~\citet{Shao2006} carried out a simulation study of the 
degree distribution in a standard mixed attachment growing network. 
More recently,~\cite{Pachon2018} investigated the scale-free 
property of the degree distribution in an analogous model through 
recursive formulations. We describe the construction of a hybrid 
random network
in Section~\ref{Sec:HRNs}, and study theoretical properties of its 
degree distributions
in Section~\ref{Sec:degdist}. We then propose estimation methods 
and explore properties of the estimators
in 
Section~\ref{Sec:estimate}, which facilitate the numerical studies 
on both synthetic datasets (cf. Section~\ref{Sec:sim}) and real 
network data (cf. Section~\ref{Sec:RDA}). 
With all results available, we also provide some interesting 
direction for future research in Section~\ref{Sec:discus}. 

\section{Hybrid Random Networks}
\label{Sec:HRNs}

\begin{figure}[ht]
	\label{fig:graphic}
	\begin{center}
		\resizebox{\textwidth}{!}{%
			\begin{tikzpicture}
				\draw[fill = gray!20] (-5, 0) ellipse (1.6 and 1) ;
				\draw (-4.2, 0) node[draw = black, circle, minimum 
				size =
				0.8cm, fill = blue!20] (i1)	{$i$} ;
				\draw (-3.5, -1.5) node[draw = black, circle, 
				minimum 
				size = 0.8cm] (u1)	{$u$} ;
				\draw[-latex, thick] (u1) -- (i1) ;
				\draw[fill = gray!20] (0, 0) ellipse (1.6 and 1) ;
				\draw (-0.8, 0) node[draw = black, circle, minimum 
				size =
				0.8cm, fill = blue!20] (i2)	{$i$} ;
				\draw (0.8, 0) node[draw = black, circle, minimum 
				size = 0.8cm, fill = blue!20] (j1)	{$j$} ;
				\draw[-latex, thick] (j1) -- (i2) ;
				\draw[fill = gray!20] (5, 0) ellipse (1.6 and 1) ;
				\draw (4.2, 0) node[draw = black, circle, minimum 
				size =
				0.8cm, fill = blue!20] (j2)	{$j$} ;
				\draw (3.5, -1.5) node[draw = black, circle, minimum 
				size = 0.8cm] (u2)	{$u$} ;
				\draw[-latex, thick] (j2) -- (u2) ;
			\end{tikzpicture}%
		}
		\caption{Three edge-addition scenarios respectively 
			corresponding to $\alpha$, $\beta$ and $\gamma$ (from 
			left to 
			right)}
	\end{center}
\end{figure}
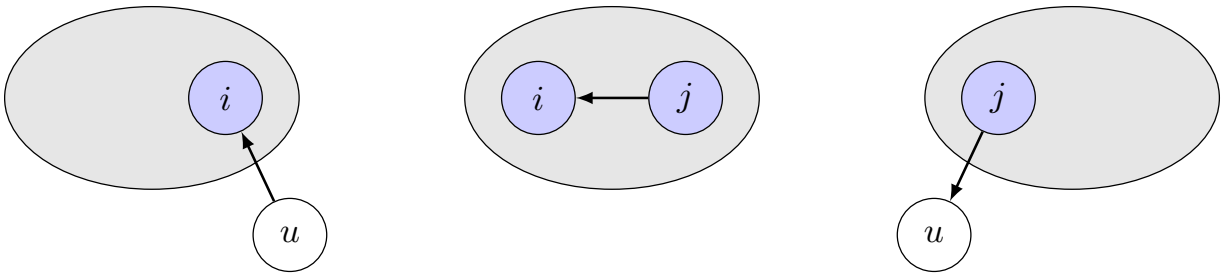

Let $\hyb_n(V_n,E_n;\alpha,\beta,\gamma,p,\deltain,\deltaout)$ 
denote the structure of a class of HRNs consisting of a vertex set 
$V_n$ and an edge set $E_n$ at time~$n$, 
parameterized by 
a set of parameters $\bftheta := (\alpha, \beta, \gamma, p, 
\deltain, \deltaout)$ subject to $\alpha + \beta + \gamma = 1$, 
$\deltain,\deltaout>0$. 
Specifically, the parameters, $\alpha$, $\beta$ and 
$\gamma$, represent the probabilities of presenting one of 
the three edge-creation scenarios at each step. With 
probability $\alpha$, there emerges a directed edge from the 
newcomer to an existing node. With probability $\gamma$, there 
emerges a directed edge from an existing node to the newcomer. With 
probability $\beta = 1 - \alpha - \gamma$, a directed edge is 
added between two existing nodes. See Figure~\ref{fig:graphic} for a 
graphical illustration.
The offset parameters $\deltain$ 
and $\deltaout$ respectively control the growth rate of in-degree 
and out-degree in the network. Another parameter $0 \le p \le 
1$ specifies the probability of executing PA when sampling the 
node(s) at the end(s) of the newly added edge at each timestamp. The 
functionality of $p$ is to balance PA and UA in the model, and 
accordingly the proposed HRN becomes more flexible than pure PA 
network model for characterizing the in-degree and out-degree tail 
distributions of real network data.

We start the network with $\hyb_0$, which is a self-looped 
single node labeled with $1$. At any subsequent point $n \ge 
1$, flip a three-sided coin, for which the probabilities of landing 
the 
three faces up are respectively $\alpha$ (associated with scenario 
1), $\beta$ (associated with scenario 2) and $\gamma$ (associated 
with scenario~3). Let $J_n \in \{1, 2, 3\}$ indicate the 
occurrence of the scenario type at time $n$, i.e. $J_n$ is a 
tri-nomial random variable on $\{1,2,3\}$ with cell probability 
$\alpha$, $\beta$ and~$\gamma$, respectively. 
The network evolves as below.
\begin{enumerate}
	\item For $J_n = 1$, we add a new 
	node $u$ to the network, connecting it to an existing node 
	$i \in V_{n - 1}$ by a directed edge pointing to $i$ with 
	probability
	\begin{equation}
		\label{eq:scenario1}
		p \times \frac{\Din_i(n - 1) + \deltain}{\sum_{k \in V_{n - 
					1}} (\Din_k(n 
			- 1) + \deltain)} + (1 - p) \times 
		\frac{1}{|V_{n - 1}|},
	\end{equation}
	where $\Din_i (n)$ is the in-degree of $i$ in $\hyb_{n}$, and 
	$|V_n|$ denotes the number of nodes at time $n$.
	\item For $J_n = 2$, we add a directed edge between two existing 
	nodes $i, j \in V_{n - 1}$, where $i$ and 
	$j$ are sampled independently. Suppose that	the newly added edge 
	is pointed (from $j$) to $i$, then the associated probability is 
	given by
	\begin{align}
		\nonumber
		&\left(p \times \frac{\Dout_j(n - 1) + \deltaout}{\sum_{k 
				\in V_{n - 1}} (\Dout_k(n - 1) + \deltaout)} + (1 - 
		p) 
		\times \frac{1}{|V_{n - 1}|}\right) 
		\\ \label{eq:scenario2} &\qquad{} \times \left(p \times 
		\frac{\Din_i(n - 1) + \deltain}{\sum_{k \in V_{n - 
					1}}(\Din_k(n - 1) + \deltain)} + (1 - p) \times 
		\frac{1}{|V_{n - 1}|}\right),
	\end{align}
	where $\Dout_i(n)$ is analogously defined as the out-degree of 
	node $i$ in~$\hyb_{n}$. Note that no new node is added to the 
	network under this scenario, hence $V_n = V_{n - 1}$. 
	Besides, there is a positive probability that a node is 
	sampled twice; If so, a self loop is created.
	\item For $J_n = 3$, a new node $u$ is appended to the network 
	by a directed edge emanating out from $j \in V_{n - 1}$ with 
	probability
	\begin{equation}
		\label{eq:scenario3}
		p \times \frac{\Dout_j(n - 1) + \deltaout}{\sum_{k \in V_{n 
					- 1}} (\Dout_k(n - 1) + \deltaout)} + (1 - p) 
		\times 
		\frac{1}{|V_{n - 1}|}.
	\end{equation} 
\end{enumerate}	

Some simplifications can be made to the conditional probabilities in 
Equations~\eqref{eq:scenario1},~\eqref{eq:scenario2} 
and~\eqref{eq:scenario3} after observing $\sum_{k \in V_{n - 1}} 
(\Din_k(n - 1) + \deltain) = n + \deltain|V_{n - 1}|$ and  
$\sum_{k \in V_{n - 1}} (\Dout_k(n - 1) + \deltaout) = n + 
\deltaout|V_{n - 1}|$ (since our initial time is $n = 0$). 
Meanwhile, the fact that the two fractions have different 
denominators in each of the conditional probabilities would have 
brought a great deal of challenges to both 
analytical computations and parameter estimations.

\section{Degree Distribution}
\label{Sec:degdist}

In this section, we investigate the in-degree and out-degree 
distributions of $\hyb_{n}$. Let $\field_n$ be the sigma field 
generated by the
evolution of a hybrid random network up to time~$n$, 
i.e., $\{\hyb_k: k\le n\}$. 
According to the evolutionary scenarios described in 
Section~\ref{Sec:HRNs}, we have for $i\in V_n$,
\begingroup
\allowdisplaybreaks
\begin{align}
	&\Prob\left(\Din_i(n + 1) - \Din_i(n) = 1 \given \field_n\right) 
	\nonumber \\
	&\qquad{} = (\alpha + \beta) \left[\frac{p\left(\Din_i(n) + 
		\deltain\right)}{\sum_{k \in V_{n}} \left(\Din_k(n) + 
		\deltain\right)} + 
	\frac{(1 - p)}{|V_{n}|}\right], \label{eq:indegconddist}
	\\ &\Prob\left(\Dout_i(n + 1) - \Dout_i(n) = 1 \given 
	\field_n\right) \nonumber \\
	&\qquad{} = (\beta + \gamma) \left[\frac{p\left(\Dout_i(n) 
		+ \deltaout \right)}{\sum_{k \in V_{n}} \left(\Dout_k(n) + 
		\deltaout \right)} + \frac{(1 - p)}{|V_{n}|}\right].
	\label{eq:outdegconddist}
\end{align}
\endgroup
We present important theoretical results on the in- and out-degree 
sequences in an HRN. Relevant proofs of the theorems in this section 
are collected in Appendices~\ref{sec:appthm1},~\ref{sec:appthm2} 
and~\ref{sec:appthm3}. 

\subsection{Expected In- and Out-Degrees}
The next theorem specifies the growth rates of 
$\Expe\left[\Din_i(n)\right]$ and $\Expe\left[\Dout_i(n)\right]$ for 
a fixed node $i \in V_n$.
\begin{theorem}
	\label{thm:focus}
	There exist $M_1, M_2 >0$ such that
	$$\sup_{i \ge 1} 
	\frac{\Expe\left[\Din_i(n)\right]}{n^{C_1}} 
	\le M_1 \qquad \mbox{and} \qquad
	\sup_{i \ge 1} \frac{\Expe\left[\Dout_i(n)\right]}{n^{C_2}} 
	\le M_2,$$
	where $C_1$ and $C_2$ are respectively given by
	$$C_1 = \frac{(\alpha + \beta)p}{1 + \deltain(1 - \beta)} \qquad 
	\mbox{and} \qquad C_2 = \frac{(\beta + \gamma)p}{1 + \deltaout(1 
		- \beta)}.$$
\end{theorem}
It is worth noting that the growth rates $C_1$ and $C_2$ are smaller 
than those in a pure directed PA model (i.e., $p = 1$). This 
suggests that incorporating a non-negligible number of 
uniformly added edges creates lighter distributional tails for both 
in- and out-degrees.

\subsection{Almost Sure Convergence of the In- and Out-Degrees}
Next, we study the asymptotic properties 
of $\Din_i(n)$ and $\Dout_i(n)$ by utilizing \emph{martingale} 
formulations~\citep[Chapter 4]{Durrett2006}. By the 
conditional probability in Equation~\eqref{eq:indegconddist}, we 
have for $i\in V_n$,
$$\Prob\left(\Din_i(n + 1) - \Din_i(n) = 1 \given \field_n\right) 
\ge \frac{p\left(\Din_i(n) + \deltain\right)(\alpha + \beta)}{1 + n 
	+ 
	\deltain|V_n|},$$
which implies that for a fixed $i$,
\begin{equation}\label{eq:Mg_in}
	M^{\rm in}_{n + 1} := \frac{\Din_i(S_i+n + 1) + 
		\deltain}{\prod_{k = 
			0}^{n} \left( 1 +
		\frac{p(\alpha + \beta)}{S_i + k+1 + 
			\deltain|V_{S_i+k}|}\right)}
\end{equation}
is a \emph{sub-martingale} with respect to the filtration 
$(\field_n)_{n\ge 0}$, 
where $S_i := \inf\{n \ge 0: |V_n| = 
i\}$. Analogously, based on 
Equation~\eqref{eq:outdegconddist}, we construct another 
sub-martingale 
sequence for out-degrees:
\begin{equation}\label{eq:Mg_out}
	M^{\rm out}_{n + 1} := \frac{\Dout_i(S_i+n + 1) + 
		\deltaout}{\prod_{k 
			= 0}^{n} \left(1 + \frac{p(\beta + \gamma)}{S_i + k +1+ 
			\deltaout|V_{S_i+k}|}\right)}.
\end{equation}
In the proof of Theorem~\ref{thm:degnum}, we specify the asymptotic 
orders of the 
denominator in Equation~\eqref{eq:Mg_in} (a similar argument also 
applies to the denominator in Equation~\eqref{eq:Mg_out}).
Then applying the martingale convergence theorem~\citep[Theorem 
4.2.11]{Durrett2006} gives the following convergence results 
for the in- and out-degrees of a fixed node.
Details of the proof of Theorem~\ref{thm:mtg} are given in 
Appendix~\ref{sec:appthm2}.
\begin{theorem}
	\label{thm:mtg}
	For a fixed node $i$, there exist finite random variables 
	$\zeta_i$ and $\xi_i$ such that as $n 
	\to \infty$,
	$$\left(\frac{\Din_i(n)}{n^{C_1}},\frac{\Dout_i(n)}{n^{C_2}}\right)
	\almostsure 
	\left(\zeta_i,\xi_i \right),$$
	where $C_1$ and $C_2$ are identical to those developed in 
	Theorem~\ref{thm:focus}.
\end{theorem}

\subsection{Expected Degree Counts}
In addition, we develop the 
asymptotics for $N_m^{\rm in}(n)/n$, and 
$N_m^{\rm out}(n)/n$, $m\ge 0$, i.e., the empirical proportional of 
nodes with in- or
out-degree $m$ 
in $\hyb_n$. 
Let ${\rm NB}(n, q)$ represent a 
negative binomial random variable with generating function
\[
(s+(1-s)/q)^{-n},\qquad s\in[0,1].
\]
\begin{theorem}
	\label{thm:degnum}
	Define $\tilde{\delta}_{\rm in} := \deltain / p + (1 - 
	p)/\bigl(p (1 - \beta)\bigr)$ and  $\tilde{\delta}_{\rm out} := 
	\deltaout / p + (1 - p)/\bigl(p (1 - \beta)\bigr)$. 
	Let
	${\rm NB}(\tdeltain, p_1)$, ${\rm NB}(1+\tdeltain, p_1)$, 
	${\rm NB}(\tilde{\delta}_{\rm out}, p_2)$ and ${\rm 
		NB}(1+\tilde{\delta}_{\rm out}, p_2)$,
	be four independent negative binomial random variables, and set 
	$T_{\rm in}$ and $T_{\rm out}$ to be two independent exponential
	random variables with rates
	\[\frac{1 + \deltain(1 - 
		\beta)}{p(\alpha + \beta)} \qquad \text{and} \qquad 
	\frac{1 + \deltaout(1 - 
		\beta)}{p(\beta + \gamma)},\]
	respectively (which are also independent from ${\rm 
		NB}(\tdeltain, p_1)$, ${\rm NB}(1+\tdeltain, p_1)$, 
	${\rm NB}(\tilde{\delta}_{\rm out}, p_2)$ and ${\rm 
		NB}(1+\tilde{\delta}_{\rm out}, p_2)$).
	As $n \to 
	\infty$, we have
	\[\frac{N_m^{\rm in}(n)}{n} \stackrel{p}{\longrightarrow} 
	\tilde{\psi}_m^{\rm in} 
	\qquad 
	\text{and} \qquad\frac{N_m^{\rm out}(n)}{n} 
	\stackrel{p}{\longrightarrow} 
	\tilde{\psi}_m^{\rm out}, \]
	where
	\begingroup
	\allowdisplaybreaks
	\begin{align}
		\tilde{\psi}_{m}^{\rm in} &= {\alpha}\,
		\Prob\left({\rm NB}\left(\tilde{\delta}_{\rm in}, 
		e^{-T_{\rm in}}\right) = m\right) \nonumber \\
		&\qquad{}+ {\gamma}
		\, \Prob\left(1 + {\rm NB}\left(1+\tilde{\delta}_{\rm in}, 
		e^{-T_{\rm in}}\right) = m\right),\label{eq:psi_in}
		\\ \tilde{\psi}_{m}^{\rm out} &= {\alpha}\,
		\Prob\left({\rm NB}\left(\tilde{\delta}_{\rm out}, 
		e^{-T_{\rm out}}\right) = m\right) 
		\nonumber \\
		&\qquad{}+ {\gamma} \,\Prob\left(1 + {\rm 
			NB}\left(1+\tilde{\delta}_{\rm 
			out}, e^{-T_{\rm out}}\right) = 
		m\right).\label{eq:psi_out}
	\end{align}
	\endgroup	
\end{theorem}
We conclude this section by remarking that the limit functions 
in Equations~\eqref{eq:psi_in} and \eqref{eq:psi_out} coincide with 
those from 
a
pure PA network with parameters 
$
(\alpha, \beta,\gamma,\tilde{\delta}_\text{in}, 
\tilde{\delta}_\text{out}).
$
In fact, when $\beta=0$, the HRN is identical to a pure PA network 
with $(\alpha, 0,\gamma,\tilde{\delta}_\text{in}, 
\tilde{\delta}_\text{out})$,
where all established results for the pure PA model can be readily 
applied.
The major goal in the proof of Theorem~\ref{thm:degnum} is to show 
that
the discrepancy caused by having random number of edges is 
negligible when $n$ is large.

\section{Parameter Estimation}
\label{Sec:estimate}

In this section, we propose our estimation scheme for the 
parameters in the HRN model described in Section~\ref{Sec:HRNs}, 
under a few regularity conditions given as follows. 
We assume the evolution history of the entire 
network is available since the beginning, recorded in the edge list 
$\bE := \{E_k\}_{k = 0}^{n - 1}$, where $E_0 = (1, 1)$ is 
deterministic. 
Notice that $\gamma = 1 - (\alpha + \beta)$ completely depends on 
$\alpha$ and $\beta$ so that the model is parametrized in terms of 
$(\alpha,\beta, p, \deltain, \deltaout)$. We also assume $0 \le p 
\le 1$, $0 \le \alpha, \beta < 1$ and $0 < \alpha + \beta \le 1$, 
where the latter two jointly ensure the exclusion of the trivial 
cases of either $\alpha$, $\beta$ or $\gamma$ taking value $1$.  The 
offset
parameters $\deltain$ and $\deltaout$ are assumed to be positive and 
finite.

\subsection{Maximum Likelihood Estimation}
\label{subsec:MLE}

In a slight abuse of notation, let $E_k = (v_{k, 1}, v_{k, 2})$ 
represent the edge (from $v_{k, 1}$ to $v_{k, 2}$) 
added at time $k$, $v_{k, 1}$ and $v_{k, 2}$ can be the nodes from 
the existing network or newcomers. According to 
Equations~\eqref{eq:scenario1},~\eqref{eq:scenario2} 
and~\eqref{eq:scenario3}, the likelihood of the model is given by
\begin{align*}
	L(\bftheta \given \bE) &= \prod_{k = 1}^{n} \left[\alpha 
	\left(\frac{p (\Din_{v_{k, 2}}(k - 1) + \deltain)}{k + 
		\deltain |V_{k - 1}|} + \frac{1 - p}{|V_{k - 
			1}|}\right)\right]^{\indicator_{\{J_k = 1\}}}
	\\ &\qquad{} \times \prod_{k = 1}^{n} \left[\beta 
	\left(\frac{p 
		(\Dout_{v_{k, 1}}(k - 1) + \deltaout)}{k+ 
		\deltaout|V_{k - 
			1}|} + \frac{1 - p}{|V_{k - 1}|}\right) \right.
	\\ &\qquad{}\qquad{} \times \left. \left(\frac{p (\Din_{v_{k, 
				2}}(k - 
		1) + \deltain)}{k + \deltain|V_{k - 1}|} + 
	\frac{1 - p}{|V_{k - 1}|}\right)\right]^{\indicator_{\{J_k = 
			2\}}}
	\\ &\qquad{} \times \prod_{k = 1}^{n} \left[(1 - \alpha - 
	\beta) 
	\left(\frac{p (\Dout_{v_{k, 1}}(k - 1) + \deltaout)}{k + 
		\deltaout|V_{k - 1}|} + \frac{1 - p}{|V_{k - 
			1}|}\right)\right]^{\indicator_{\{J_k = 3\}}},
\end{align*}
then the log-likelihood becomes
\begingroup
\allowdisplaybreaks
\begin{align*}
	&\log L(\bftheta \given \bE) 
	\\ &\quad{}= \log \alpha \sum_{k = 1}^{n} 
	\indicator_{\{J_k = 1\}} + \log \beta \sum_{k = 1}^{n - 1} 
	\indicator_{\{J_k = 2\}} + \log (1 - \alpha - \beta) \sum_{k = 
		2}^{n} \indicator_{\{J_k = 3\}}
	\\ &\qquad{} + \sum_{k = 1}^{n} \log \left[\bigl(p \Din_{v_{k, 
			2}}(k 
	- 1) + \deltain \bigr)|V_{k - 1}| + (1 - p)k\right] 
	\indicator_{\{J_k = \{1, 2\}\}}
	\\ &\qquad{} + \sum_{k = 1}^{n} \log \left[\bigl(p \Dout_{v_{k, 
			1}}(k 
	- 1) + \deltaout \bigr)|V_{k - 1}| + (1 - p)k\right] 
	\indicator_{\{J_k = \{2, 3\}\}}
	\\ &\qquad{} - \sum_{k = 1}^{n} \log \left[k|V_{k - 
		1}| + |V_{k - 1}|^2 \deltain\right] \indicator_{\{J_k = \{1, 
		2\}\}}
	\\ &\qquad{} - \sum_{k = 1}^{n} \log \left[k|V_{k - 
		1}| + 
	|V_{k - 1}|^2 \deltaout\right] \indicator_{\{J_k = \{2, 3\}\}},
\end{align*}
\endgroup
from which we see that the score functions of $\alpha$ and $\beta$ 
are 
independent of those of the other parameters. 

Carrying out an analogous 
analysis as in~\cite[Section 3.1]{Wan2017}, we find that the MLEs
for $\alpha$ and $\beta$ are respectively given by
\[\hat{\alpha}^{\, \rm MLE} = \frac{1}{n} \sum_{k = 1}^{n} 
\indicator_{\{J_k = 
	1\}} \qquad \text{and} \qquad 
\hat{\beta}^{\, \rm MLE} = \frac{1}{n} \sum_{k = 1}^{n} 
\indicator_{\{J_k 
	= 2\}}.\]
To develop the MLEs for $\deltain$, $\deltaout$ and $p$, we 
calculate their score functions to get
\begin{align}
	\frac{\partial}{\partial \deltain} \log L(\bftheta \given 
	\bE) 
	&= \sum_{k = 1}^{n} \frac{|V_{k - 1}|}{\bigl(p \Din_{v_{k, 
				2}}(k - 1) + \deltain \bigr)|V_{k - 1}| + (1 - p)k} 
	\, \indicator_{\{J_k = \{1, 2\}\}}
	\nonumber\\ &\qquad{}- \sum_{k = 1}^{n} \frac{|V_{k - 1}|}{k + 
	|V_{k - 
			1}| \deltain} \, \indicator_{\{J_k = 
		\{1, 2\}\}},\label{eq:score_din}
	\\ 	\frac{\partial}{\partial \deltaout} \log L(\bftheta \given 
	\bE) 
	&= \sum_{k = 1}^{n} \frac{|V_{k - 1}|}{\bigl(p \Dout_{v_{k, 
				1}}(k - 1) + \deltaout \bigr)|V_{k - 1}| + (1 - p)k} 
	\, \indicator_{\{J_k = \{2, 3\}\}}
	\nonumber\\ &\qquad{}- \sum_{k = 1}^{n} \frac{|V_{k - 1}|}{k + 
	|V_{k - 
			1}| \deltaout} \, \indicator_{\{J_k = 
		\{2, 3\}\}},\label{eq:score_dout}
	\\ \frac{\partial}{\partial p} \log L(\bftheta \given \bE) 
	&= \sum_{k = 1}^{n} \frac{\bigl(\Din_{v_{k, 2}}(k - 1)|V_{k - 
			1}| - 
		k\bigr)\indicator_{\{J_k = \{1, 2\}\}}}{\bigl(p \Din_{v_{k, 
				2}}(k - 
		1) + \deltain \bigr)|V_{k - 1}| + (1 
		- p)k}
	\nonumber\\ &\qquad{}+ \sum_{k = 1}^{n} \frac{\bigl(\Dout_{v_{k, 
	1}}(k - 
		1)|V_{k - 
			1}| - k\bigr)\indicator_{\{J_k = \{2, 3\}\}}}{\bigl(p 
		\Dout_{v_{k, 1}}(k - 1) + \deltaout \bigr)|V_{k - 
			1}| + (1 - p)k}.\label{eq:score_p}
\end{align}

We proceed by first setting \eqref{eq:score_din} to 0. Note that due 
to the 
randomness of $|V_{k-1}|$, the methodology given in \cite{Wan2017} 
is not 
directly applicable. Instead, we approximate the score function 
\eqref{eq:score_din}
as follows.
\begin{align*}
	\sum_{k = 1}^{n} &\frac{|V_{k - 1}|}{\bigl(p \Din_{v_{k, 
				2}}(k - 1) + \deltain \bigr)|V_{k - 1}| + (1 - p)k} 
	\, \indicator_{\{J_k = \{1, 2\}\}}\\
	&= \sum_{k = 1}^{n} \frac{\indicator_{\{J_k = \{1, 
	2\}\}}}{\bigl(p \Din_{v_{k, 
				2}}(k - 1) + \deltain \bigr) + (1 - p)k/|V_{k - 
				1}|}\\ 
	&= \frac{1}{p}\sum_{k = 1}^{n} \frac{\indicator_{\{J_k = \{1, 
	2\}\}}}{\Din_{v_{k, 
				2}}(k - 1) + \tilde{\delta}_\text{in}} + 
				R_\text{in}(n),						
\end{align*}
where
\begin{align*}
	&R_\text{in}(n) =\\
	&\sum_{k = 1}^{n} \frac{\indicator_{\{J_k = \{1, 2\}\}}}{p}\left(
	\frac{1}{\Din_{v_{k,2}}(k-1)+\deltain/p+(1-p)k/|V_{k-1}|}
	-\frac{1}{\Din_{v_{k, 
				2}}(k - 1) + \tilde{\delta}_\text{in}}\right).
\end{align*}
Therefore,
\begin{align*}
	|R_\text{in}(n)|&\le \frac{1}{p}\sum_{k=1}^n 
	\frac{(1-p)/p\left|k/|V_{k-1}|-1/(1-\beta)\right|\indicator_{\{J_k
	 = \{1, 
	2\}\}}}{(\Din_{v_{k,2}}(k-1)+\deltain/p+(1-p)k/|V_{k-1}|)(\Din_{v_{k,2}}(k
	 - 1) + \tilde{\delta}_\text{in})}\\
	&\le \frac{1-p}{p^2}\sum_{k=1}^n 
	\frac{|k/|V_{k-1}|-1/(1-\beta)|}{(\deltain/p+(1-p)k/|V_{k-1}|)\tilde{\delta}_\text{in}}.
\end{align*}
Since $|V_{n-1}|/n\convas 1/(1-\beta)$, then by the Ces\`aro 
convergence of
random variables, we have $|R_\text{in}(n)|/n\convas 0$.
Then the approximate score equation in \eqref{eq:score_din} becomes
\begin{align*}
	\frac{1}{n}\sum_{k=1}^n \frac{\indicator_{\{J_k = \{1, 
	2\}\}}}{\Din_{v_{k, 
				2}}(k - 1) + \tilde{\delta}_\text{in}}
	&= \frac{1}{n}\sum_{k = 1}^{n} \frac{|V_{k - 1}|}{k + |V_{k - 
			1}| \deltain} \, \indicator_{\{J_k = 
		\{1, 2\}\}}.
\end{align*}
Applying the method in \cite{Wan2017} further yields the following 
approximate score function:
\begin{align}\label{eq:score_din_approx}
	\sum_{m=0}^\infty\frac{N^\text{in}_{>m}(n)/n}{m+\tilde{\delta}_\text{in}}
	&= \frac{\gamma}{\tilde{\delta}_\text{in}} + 
	\frac{(\alpha+\beta)(1-\beta)}{1+\deltain(1-\beta)},
\end{align}
where $N^\text{in}_{>m}(n)$ denotes the number of nodes with
in-degree strictly greater than $m$ in $\hyb_n$.

Similarly, the score equation with respect to \eqref{eq:score_dout} 
can be approximated by 
\begin{align}\label{eq:score_dout_approx}
	\sum_{m=0}^\infty\frac{N^\text{out}_{>m}(n)/n}{m+\tilde{\delta}_\text{out}}
	&= \frac{\alpha}{\tilde{\delta}_\text{out}} + 
	\frac{(\beta+\gamma)(1-\beta)}{1+\deltaout(1-\beta)},
\end{align}
with $N^\text{out}_{>m}(n)$ being the number of nodes with
out-degree strictly greater than $m$ in $\hyb_n$.
However, with \eqref{eq:score_din_approx} and 
\eqref{eq:score_dout_approx} 
available, the approximation to the third score equation in 
\eqref{eq:score_p}
leads to a deterministic solution of $p=1$. This indicates standard 
approaches to find
MLE as in \cite{Wan2017} are not able to give us the desirable 
results.
Alternatively, in the next section, we formulate the MLE searching 
procedure as a nonlinear optimization
problem with constraint $p\in [0,1]$.





\subsection{Nonlinear optimization}


Having seen that standard ways to find MLE by solving score equations
do not produce reasonable estimates, 
we consider the MLE-searching procedure as a
nonlinear optimization problem with properly identified constraints 
such 
that $\alpha,\beta,p\in[0,1]$ and $\deltain,\deltaout >0$.
Specifically, we adopt the \emph{Nelder-Mead} 
(N-M) algorithm proposed by~\citet{Nelder1965}. The N-M algorithm is 
appealing for efficiency as it usually converges within a relatively 
small number of iterations. Despite limited knowledge about the 
theoretical results of the N-M algorithm~\citep{Lagarias1998}, its 
utilization is widespread in the community since it generally 
performs well in practice. One practical issue of the 
algorithm is that its convergence is quite 
sensitive to the choice of the initial simplex. An improper initial 
simplex has become the main cause of the algorithm breakdown.  

Having this in mind, we 
back up with an alternative --- a \emph{Bayesian} estimation based 
on 
\emph{Markov chain Monte Carlo} (MCMC) algorithms. Specifically, we 
consider a \emph{Metropolis-Hastings} (M-H)
algorithm~\citep{Metropolis1953, Hastings1970}. Being a classical 
approach, the fundamentals of the M-H algorithms have been 
extensively 
elaborated in a wide range of texts, such as~\citet{Chen2010, 
	Liang2010, Gelman2013}. Here we only present a few essential 
steps. 

Let $\pi(\bftheta ; \bfpsi)$ be the prior distribution of 
$\bftheta$, where $\bfpsi$ is a collection of hyper-parameters. 
Under this setting, the likelihood function $L(\bftheta \given 
\bE)$ is the posterior distribution of $\bftheta$. As it is 
difficult to sample $\bftheta$ from this posterior distribution 
directly, the M-H algorithm generates a Markov 
process whose stationary distribution is the same as the 
posterior distribution, and the generation of the process is done 
in an iterative manner. Let $\bftheta^{(t)}$ be the 
estimates from the $t$-th iteration, and let $Q(\bftheta^{\rm prop} 
\given \bftheta^{(t)})$ denote the proposal density governing the 
transition probability from the current estimates to a proposed set 
of candidates. Suppose that the distribution $Q$ is symmetric, then 
the 
acceptance rate $a(\bm{\theta}^{\rm prop} \given \bm{\theta}^{(t)})$ 
is given by
\begin{align*}
	a(\bftheta^{\rm prop} \given \bftheta^{(t)}) &= 
	\min\left\{\frac{L(\bftheta^{\rm prop} \given \bE) 
		Q(\bftheta^{\rm prop} \given 
		\bftheta^{(t)})}{L(\bftheta^{(t)} 
		\given \bE) 
		Q(\bftheta^{(t)} \given \bftheta^{\rm prop})}, 1\right\} 
	\\ &= 
	\min\left\{\frac{L(\bftheta^{\rm prop} \given 
		\bE)}{L(\bftheta^{(t)} 
		\given \bE)}, 1\right\}.
\end{align*}
This can be done by generating a standard uniform random variable 
$U$ such that $\bftheta^{(t + 1)} = \bftheta^{\rm prop}$ if $U < 
a(\bftheta^{\rm prop} \given \bftheta^{(t)})$; $\bftheta^{(t + 1)} = 
\bftheta^{(t)}$, otherwise. 

There are multiple ways of selecting an 
appropriate proposal distribution $Q(\bftheta^{\rm prop} 
\given \bftheta^{(t)})$, where a simple approach based 
on random walk is adopted. 
Consider $\bftheta^{\rm prop} = 
\bftheta^{(t)} + \bfepsilon$, where $\bfepsilon$ is a random 
variable for step size. Note that the conditional 
distribution $Q(\bftheta^{\rm prop} 
\given \bftheta^{(t)})$ is fully specified by the distribution of 
$\bfepsilon$. Thus, we only need 
to choose a symmetric distribution for $\bfepsilon$ to satisfy the 
required condition, such as normal or uniform distribution. One 
drawback of MCMC 
algorithms is the lack of theoretical foundation for the assessment 
of convergence. Since the initial samples of~$\bftheta$ from the 
prior 
distribution may fall into a low density of the target posterior 
distribution, a sufficiently large \emph{burn-in} period is always 
necessary. The number of iterations needed for the algorithm to 
converge is closely related to its convergence 
rate~\citep{Mengersen1996}, which is practically unwieldy in 
general. Here we will rely on a few widely-accepted graphical 
diagnostics to assess the convergence of MCMC algorithms, 
such as time-series plots and running mean plots~\citep{Smith2007}.

\section{Simulations}
\label{Sec:sim}

In this section, we carry out an extensive simulation study 
along with a sensitivity analysis for the 
estimation of the parameters for HRNs. We focus on the 
performance of the N-M and M-H algorithms under different 
combinations
of $(\alpha,\beta,p)$. Specifically, we consider $p \in \{0.8, 
0.6, 0.2\}$ (respectively corresponding to dominant PA, roughly even 
PA and UA and dominant UA) paired with $(\alpha, 
\beta) \in \{(0.8, 0.1), (0.45, 0.1), (0.1, 0.8)\}$. The other two 
offset parameters are set to be $\deltain = 1.3$ and $\deltaout = 
0.7$.

For each setting, we generate $R = 100$ replicates of independent 
HRNs with $n=10,000$ edges. When applying the M-H algorithm, we use 
non-informative priors. The burn-in number is set to be 10,000, and 
the number of 
iterations after burn-in is 20,000. To avoid auto-correlation 
in the posterior sample, a \emph{thinning} sampling of gap 500 is 
used. In Tables~\ref{tab:simp08},~\ref{tab:simp06} 
and~\ref{tab:simp02}, we present the point estimates, the absolute 
percentages of bias and the standard errors based on the 
simulation results.

\begin{table}[tbp]
	\caption{Simulation results with large $p = 0.8$.}
	\label{tab:simp08}\par
	\resizebox{\linewidth}{!}{
		\begin{tabular}{cccccccccccc} 
			\toprule
			&
			& \multicolumn{5}{c}{Nelder-Mead Algorithm}
			& \multicolumn{5}{c}{Metropolis-Hastings Algorithm}
			\\ 
			\cmidrule(lr){3-7} \cmidrule(lr){8-12}
			\multicolumn{2}{c}{Parameters} & $\hat{\alpha}$ & 
			$\hat{\beta}$ & $\hat{p}$ & $\hat{\delta}_{\rm in}$ & 
			$\hat{\delta}_{\rm out}$ & $\hat{\alpha}$ & 
			$\hat{\beta}$ & 
			$\hat{p}$ & $\hat{\delta}_{\rm in}$ & 
			$\hat{\delta}_{\rm out}$ \\ 
			\midrule
			$\alpha = 0.1$ & Est. & 0.0996 & 0.8004 & 0.7908 & 
			1.2304 & 0.6345 & 0.0997 & 0.8002 & 0.8183 & 1.4513 & 
			0.8331 \\
			$\beta = 0.8$ & Bias(\%) & 0.4068 & 0.0534 & 1.1653 & 
			5.6584 & 10.3172 & 0.2669 & 0.0201 & 2.2357 & 10.4235 & 
			15.9719 \\
			$p = 0.8$ & S.E. & 0.0003 & 0.0004 & 0.0010 & 0.0103 & 
			0.0058 &  0.0003 & 0.0004 & 0.0079 & 0.0640 & 0.0567 \\
			\midrule
			$\alpha = 0.8$ & Est. & 0.8003 & 0.0999 & 0.7611 & 
			1.1776 
			& 0.6251 & 0.8001 & 0.1000 & 0.8128 & 1.3351 & 0.8302 
			\\
			$\beta = 0.1$ & Bias(\%) & 0.0359 & 0.1087 & 5.1091 & 
			10.3971 & 11.9915 & 0.0118 & 0.0231 & 1.5728 & 2.6264 & 
			15.6826 \\
			$p = 0.8$ & S.E. & 0.0004 & 0.0003 & 0.0048 & 0.0154 & 
			0.0138 & 0.0004 & 0.0003 & 0.0118 & 0.0361 & 0.0449 \\
			\midrule
			$\alpha = 0.45$ & Est. & 0.4497 & 0.1003 & 0.8153& 
			1.3608 & 0.7339 & 0.4494 & 0.1005 & 0.8305 & 1.4193 & 
			0.7714 \\
			$\beta = 0.1$ & Bias(\%) & 0.0763 & 0.3218 & 1.8762 & 
			4.6677 & 4.6199 & 0.1397 & 0.4989 & 3.6774 & 8.4022 & 
			9.2520 \\
			$p = 0.8$ & S.E. & 0.0005 & 0.0003 & 0.0049 & 0.0156 & 
			0.0113 & 0.0005 & 0.0003 & 0.0098 & 0.0324 & 0.0236  \\
			\bottomrule
	\end{tabular}}
\end{table}

\begin{table}[tbp] 
	\caption{Simulation results with moderate $p = 0.6$.}
	\label{tab:simp06}\par
	\resizebox{\linewidth}{!}{
		\begin{tabular}{cccccccccccc} 
			\toprule
			&
			& \multicolumn{5}{c}{Nelder-Mead Algorithm}
			& \multicolumn{5}{c}{Metropolis-Hastings Algorithm}
			\\ 
			\cmidrule(lr){3-7} \cmidrule(lr){8-12}
			\multicolumn{2}{c}{Parameters} & $\hat{\alpha}$ & 
			$\hat{\beta}$ & $\hat{p}$ & $\hat{\delta}_{\rm in}$ & 
			$\hat{\delta}_{\rm out}$ & $\hat{\alpha}$ & 
			$\hat{\beta}$ & 
			$\hat{p}$ & $\hat{\delta}_{\rm in}$ & 
			$\hat{\delta}_{\rm out}$ \\ 
			\midrule
			$\alpha = 0.1$ & Est. & 0.1002 & 0.7996 & 0.6001 & 
			1.3340 & 0.7012 & 0.1003 & 0.7995 & 0.6216 & 1.5576 & 
			0.9057 \\
			$\beta = 0.8$ & Bias(\%) & 0.2090 & 0.0529 & 0.0101 & 
			2.5520 & 0.1694 & 0.3223 & 0.0645 & 3.4725 & 16.5359 & 
			22.7067 \\
			$p = 0.6$ & S.E. & 0.0003 & 0.0004 & 0.0016 & 0.0162 & 
			0.0134 & 0.0003 & 0.0005 & 0.0077 & 0.0819 & 0.0704 \\
			\midrule
			$\alpha = 0.8$ & Est. & 0.8002 & 0.1001 & 0.5713 & 
			1.1725 & 0.6319 & 0.8000 & 0.1001 & 0.6537 & 1.4981 & 
			1.0724 \\
			$\beta = 0.1$ & Bias(\%) & 0.0288 & 0.0539 & 5.0156 & 
			10.8786 & 10.7706 & 0.0055 & 0.1458 & 8.2084 & 13.2215 & 
			34.7240 \\
			$p = 0.6$ & S.E. & 0.0004 & 0.0003 & 0.0034 & 0.0115 & 
			0.0146 & 0.0004 & 0.0003 & 0.0144 & 0.0564 & 0.0752 \\
			\midrule
			$\alpha = 0.45$ & Est. & 0.4504 & 0.1001 & 0.6227 & 
			1.4097 & 0.7746 & 0.4503 & 0.1002 & 0.6643 & 1.5867 & 
			0.9064 \\
			$\beta = 0.1$ & Bias(\%) & 0.0930 & 0.1252 & 3.6493 & 
			7.7825 & 9.6363 & 0.0747 & 0.2405 & 9.6789 & 18.0667 & 
			22.7719\\
			$p = 0.6$ & S.E. & 0.0005 & 0.0003 & 0.0034 & 0.0147 & 
			0.0104 & 0.0005 & 0.0003 & 0.0154 & 0.0637 & 0.0468 \\
			\bottomrule
	\end{tabular}}
\end{table}

\begin{table}[tbp] 
	\caption{Simulation results with small $p = 0.2$.}
	\label{tab:simp02}\par
	\resizebox{\linewidth}{!}{
		\begin{tabular}{cccccccccccc} 
			\toprule
			&
			& \multicolumn{5}{c}{Nelder-Mead Algorithm}
			& \multicolumn{5}{c}{Metropolis-Hastings Algorithm}
			\\ 
			\cmidrule(lr){3-7} \cmidrule(lr){8-12}
			\multicolumn{2}{c}{Parameters} & $\hat{\alpha}$ & 
			$\hat{\beta}$ & $\hat{p}$ & $\hat{\delta}_{\rm in}$ & 
			$\hat{\delta}_{\rm out}$ & $\hat{\alpha}$ & 
			$\hat{\beta}$ & 
			$\hat{p}$ & $\hat{\delta}_{\rm in}$ & 
			$\hat{\delta}_{\rm out}$ \\ 
			\midrule
			$\alpha = 0.1$ & Est. & 0.1003 & 0.7998 & 0.2021 & 
			1.2424 & 0.8122 & 0.1004 & 0.7996 & 0.2079 & 1.4151 & 
			1.0505 \\
			$\beta = 0.8$ & Bias(\%) & 0.2895 & 0.0259 & 1.0335 & 
			4.6396 & 13.8146 & 0.3811 & 0.0503 & 3.8097 & 8.1307 & 
			33.3683 \\
			$p = 0.2$ & S.E. & 0.0003 & 0.0004 & 0.0010 & 0.0249 & 
			0.0268 & 0.0003 & 0.0004 & 0.0023 & 0.0788 & 0.0750 \\
			\midrule
			$\alpha = 0.8$ & Est. & 0.7993 & 0.1005 & 0.1971 & 
			1.2281 & 0.8923 & 0.7992 & 0.1006 & 0.2321 & 1.6970 & 
			1.2663 \\
			$\beta = 0.1$ & Bias(\%) & 0.0863 & 0.4633 & 1.4966 & 
			5.8588 & 21.5482 & 0.1052 & 0.5901 & 13.8237 & 23.3961 & 
			44.7213 \\
			$p = 0.2$ & S.E. & 0.0004 & 0.0003 & 0.0039 & 0.0410 & 
			0.0436 & 0.0004 & 0.0003 & 0.0071 & 0.0902 & 0.0996 \\
			\midrule
			$\alpha = 0.45$ & Est. & 0.4495 & 0.0998 & 0.2125 & 
			1.4816 & 0.8271 & 0.4493 & 0.0999 & 0.2398 & 1.7302 & 
			1.2277\\
			$\beta = 0.1$ & Bias(\%) & 0.1186 & 0.1863 & 5.8757 & 
			12.2570 & 15.3696 & 0.1644 & 0.1022 & 16.6028 & 24.8628 
			& 42.9845\\
			$p = 0.2$ & S.E. & 0.0005 & 0.0003 & 0.0026 & 0.0265 & 
			0.0249 & 0.0005 & 0.0003 & 0.0072 & 0.0703 & 0.1108\\
			\bottomrule
	\end{tabular}}
\end{table}

Overall, both algorithms provide estimates with low bias for 
$(\alpha,\beta,p)$, across all combinations of the parameters,
but we do observe that the N-M algorithm outperforms the M-H 
algorithm.

In particular, for $p$ which controls the percentage of 
edges produced by the PA rule, 
the N-M algorithm is preferred as the standard 
errors are consistently smaller. 
Estimates for $\deltain$ 
and $\deltaout$ from both algorithms tend to be biased, especially 
when $p$ is small (cf.\ Table~\ref{tab:simp02}; when the UA part 
dominates), though the estimated $\deltain$ 
is slightly less biased than the estimated $\deltaout$. 
When $p=0.8$, Table \ref{tab:simp08} reveals that
the N-M method produces the best (small bias and small standard 
errors)
estimated $(\deltain, \deltaout)$ for the combination
$(\alpha,\beta,\gamma) = (0.45,0.1,0.45)$.
When $p=0.2$ or 0.6, from Tables \ref{tab:simp06} and 
\ref{tab:simp02}, we see the most accurately estimated $(\deltain, 
\deltaout)$ appear in case
$(\alpha,\beta,\gamma) = (0.1,0.8,0.1)$.
Especially for $p = 0.6$, the N-M algorithm 
provides the most accurate estimation overall. 

\begin{figure}[tbp]
	\centering
	\includegraphics[width=\textwidth]{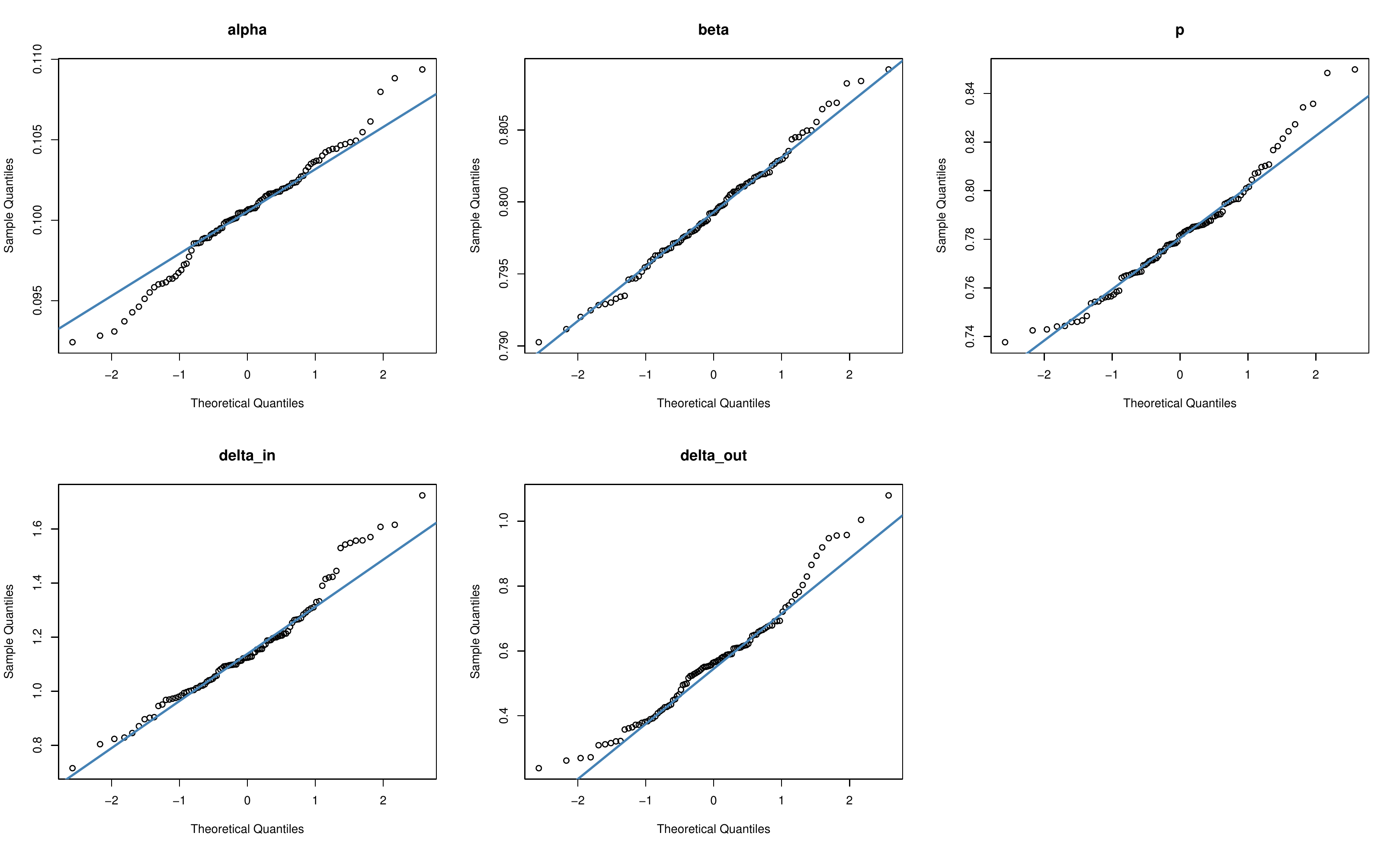}
	\caption{Q-Q plots for the estimates based on $100$ 
		independently simulated HRNs with $\alpha = 0.1$, $\beta = 
		0.8$, 
		$p = 0.8$, $\deltain = 1.3$ and $\deltaout = 0.7$.}
	\label{fig:qq}
\end{figure}

The simulation results reveal that $\hat{\alpha}$, $\hat{\beta}$ and 
$\hat{p}$ are all unbiased from both algorithms, and there is no 
significant difference in {\em relative efficiency} observed for 
estimating the scenario probabilities $\alpha$ and $\beta$. However, 
there is a noticeable efficiency gain in estimating the 
tuning parameter $p$ by using the N-M algorithm, rendering it a 
preferred approach. Besides, we look into the distributions of the 
estimates. The standard central limit theorem ensures that the 
limiting distributions of $\hat{\alpha}^{\, \rm MLE}$ and 
$\hat{\beta}^{\,\rm MLE}$ are normal. However, since the score 
functions for $\delta_{\rm in}$, $\delta_{\rm out}$ and $p$ are not 
separable, no standard approach can be readily used to uncover their 
limiting distributions. Nonetheless, the approximation method 
developed in Theorem~\ref{thm:degnum} plausibly suggests that they 
may follow a Gaussian law asymptotically as well. To verify, we show 
the quantile-quantile (Q-Q) plots for the estimates from the case of 
$\alpha = 0.1$, $\beta = 0.8$ and $p = 0.8$ as an example; see 
Figure~\ref{fig:qq}. The Q-Q plots imply that each of the estimates 
seems to follow a normal distribution marginally. We then run 
analogous analyses on the simulated HRNs under different parameter 
settings, and obtain the same pattern. So, those Q-Q plots are not 
repeatedly presented. 

Albeit the relatively better performance of the N-M method, the 
algorithm, as mentioned, undergoes the limitation of sensitivity to 
the initial value. Under the current setting, the N-M algorithm is 
able to provide estimation results over 90\% of the simulation runs 
with a fixed initial start given by $\alpha^{(0)} = \beta^{(0)} = 
\gamma^{(0)} = 1/3$, $p^{(0)} = 1/2$ and $\deltain^{(0)} = 
\deltaout^{(0)} = 1$. When we use a random initial start (e.g., 
spacing $\alpha$, $\beta$ and $\gamma$ randomly on the unit 
interval, sampling $p$ from a standard uniform distribution and 
sampling $\deltain$ 
and $\deltaout$ independently from a standard exponential 
distribution), the success rate may drop significantly to $60\%$ 
or less. The failure of the algorithm is primarily due to the 
inaccurate start of $\deltain$ and $\deltaout$. We have also run 
some experiments on smaller 
networks. When reducing the simulated network size to 5,000, 
the success rate of the N-M algorithm declines to 35\% or less 
even with a fixed initial simplex. 

In contrast, the M-H 
algorithm is more robust, as it is always able to produce estimation 
results regardless of the size of the network. Specifically, we 
consider a 
random spacing of $\alpha$, $\beta$ and $\gamma$ on the unit 
interval, sample $\deltain$ and $\deltaout$ independently from a 
standard exponential distribution, and let $p$ start from a value 
close to $1$ (e.g., $1 - 10^{-4}$), which indicates an almost 
perfect linear PA. Simulation results show that the choice 
of a non-informative prior of $p$ 
(i.e., sampling it uniformly from $[0, 1]$) has negligible impact on 
the final estimation results. Though we observe bias in the 
estimates of
$\deltain$ and $\deltaout$, the 100\% success rate renders the M-H 
method a competitive alternative. 

To summarize, we recommend the N-M algorithm for parameter 
estimation if there is auxiliary information available 
to decide reasonable initial values. In addition, the M-H 
algorithm is a possible backup when the N-M algorithm fails. In 
practice, we may consider an integration of the two algorithms. 
Although the 
estimates of $\deltain$ and $\deltaout$ from the M-H algorithm
may not be very accurate, they are close enough to true values 
allowing for a successful implementation of the N-M algorithm. 
Therefore, we 
may first adopt the M-H algorithm to get coarse estimates of the 
model parameters, and use them as initial values for the N-M 
algorithm, which ultimately leads to finer estimation results. It is 
worth mentioning that this initial value selection procedure 
does not actually affect the estimation results by the N-M 
algorithm, but effectively increases the probability of the 
successful implementation of the N-M algorithm. If the target 
parameter is~$p$ only but not $\deltain$ or $\deltaout$, the 
estimation results from the M-H algorithm may have been acceptable.

\section{Real Data Analysis}
\label{Sec:RDA}

In this section, we fit the proposed HRN model to two real network 
datasets: 
the Dutch Wikipedia talk network and the Facebook wall posts, both 
of which
are retrieved from 
the KONECT network data repository (\url{http://konect.cc/}). 
By investigating the timestamp information from both datasets, we 
see the existence of
two additional edge creation scenarios at each step:
\begin{enumerate}
	\item Set $J_n = 4$ (with probability $0 < \xi < 1$) if a new 
	node 
	with a 
	self-loop is added to the 
	network;
	\item Set $J_n = 5$ (with probability $0 < \eta < 1$) if two new 
	nodes 
	with a directed edge 
	connecting them are added to the network.
\end{enumerate}
Note that these two additional scenarios require minor modifications 
of the log-likelihood given in Section~\ref{Sec:estimate}, but 
they do not impose direct effect on the score functions of $p$, 
$\deltain$ and $\deltaout$. The MLEs of $\xi$ and $\eta$ are 
straightforward:
\[\hat{\xi}^{\, \rm MLE} = \frac{1}{n} \sum_{k = 1}^{n} 
\indicator_{\{J_k = 
	4\}} \qquad \text{and} \qquad 
\hat{\eta}^{\, \rm MLE} = \frac{1}{n} \sum_{k = 1}^{n} 
\indicator_{\{J_k 
	= 5\}}.\] 

\subsection{Dutch Wikipedia Talk}
\label{Sec:wiki}

In the Dutch Wikipedia talk dataset, each single node represents a 
specific user of
the Dutch Wikipedia, and the creation of a directed edge from 
node~$u$ to node $v$ 
refers to the event that user $u$ leaves 
a message on user $v$'s talk page. 
The dataset includes the communication among the users of Dutch
Wikipedia talk pages from 10/18/2002 and 
11/23/2015, consisting of three columns.
The first two columns represent users' ID and the third column gives 
a UNIX timestamp with
the time of a message posted on one's Wikipedia talk page.
For each row, the first user writes a message on the talk page of 
the second user at a timestamp given in the third column.

\begin{figure}[t!]
	\centering
	\includegraphics[width=\textwidth]{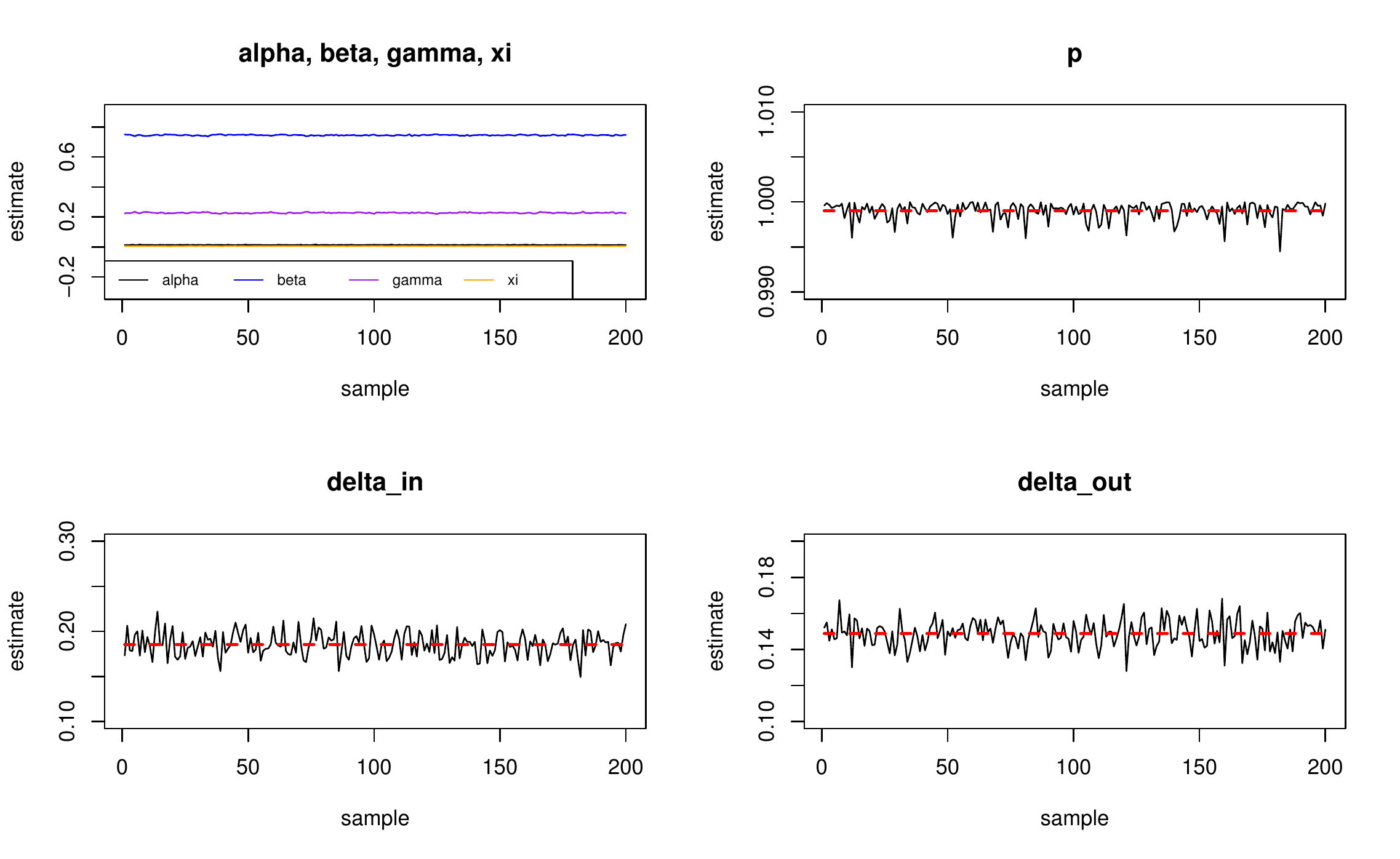}
	\caption{Graphical diagnostics of the convergence for the M-H 
		algorithm applied to the sub-network of Dutch Wikipedia talk 
		dataset; ; red dashed lines represent the 
		estimated values.}
	\label{fig:MH_wiki}
\end{figure}

We select the sub-network according to the timestamp information 
from 01/01/2013 to 03/31/2013. The sub-network is directed, 
consisting of 3,288 nodes and 21,724 directed edges. We fit the 
proposed hybrid model to the data by using the integration of the 
M-H algorithm and the N-M algorithm. For the M-H algorithm, the 
burn-in number and iteration sample size are both set at 100,000, 
and the gap for thinning sampling is 500. The convergence of the 
estimates is checked via the time-series plots in 
Figure~\ref{fig:MH_wiki}.
Using the estimates from the M-H algorithm as the initial values for 
the N-M algorithm, we get almost identical estimates given by
\[\widehat{\boldsymbol{\theta}}_{\rm WK}:=(\hat{\alpha}, 
\hat{\beta}, \hat{\gamma}, \hat{\xi}, \hat{p}, 
\hat{\delta}_{\rm in}, \hat{\delta}_{\rm out}) = (0.014, 0.745, 
0.227, 0.006, 0.999, 0.186, 0.149),\]
where the value of $\hat{p}$ is extremely close to~$1$.
The large $\hat{p}$ suggests PA dominates the evolutionary process, 
thus
leading to little difference between the proposed HRN and that 
proposed in~\citet{Wan2017}. For comparison, we compute the MLEs 
from 
the pure PA network model to get
\[\widehat{\boldsymbol{\theta}}_{\rm WK}^{\, \rm 
	(PA)}:=(\hat{\alpha}, 
\hat{\beta}, \hat{\gamma}, \hat{\xi}, \hat{\delta}_{\rm in}, 
\hat{\delta}_{\rm out}) = (0.012, 0.784, 
0.189, 0.006, 0.156, 0.152),\]
from which we see the estimates from the two models are close 
to each other.

\begin{figure}[t!]
	\centering
	\includegraphics[width=\textwidth]{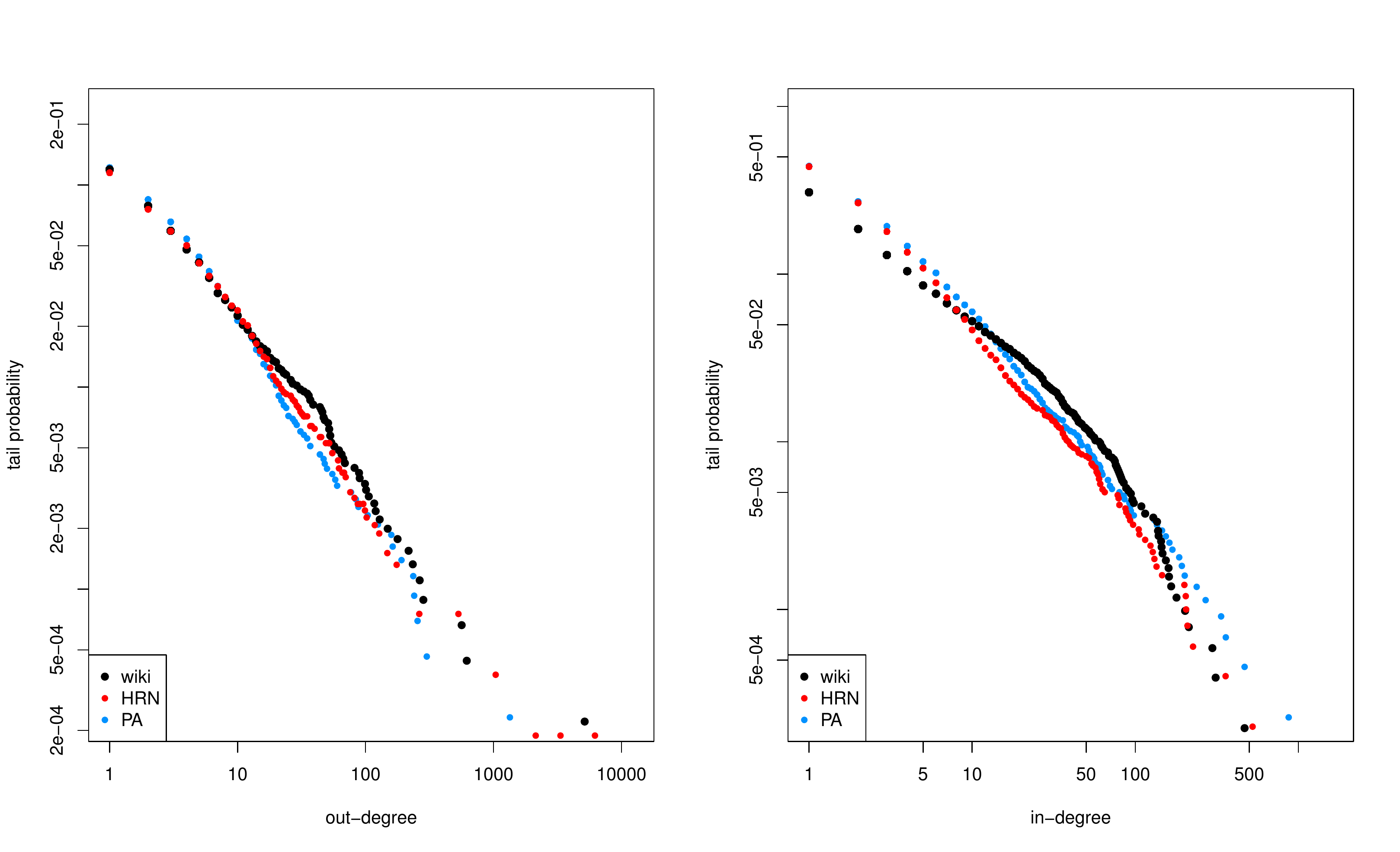}
	\caption{Out-degree and in-degree tail distributions of HRNs 
		(red), directed PA networks (blue) and the sub-network of 
		Dutch Wikipedia talk dataset (black).}
	\label{fig:tail_wiki}
\end{figure}

Next, we simulate an HRN and a pure PA network with the 
estimated parameters, and plot their corresponding empirical tail 
distributions of the out- and 
in-degrees in Figure~\ref{fig:tail_wiki}. As 
expected, the empirical out-degree and in-degree tail distributions 
from the two simulated networks are alike. It seems that the 
out-degree tail distribution (of the simulated networks) fits 
the real data better than the in-degree tail 
distribution, but the discrepancy between the in-degree tail 
distributions is also small.

\begin{figure}[t!]
	\centering
	\includegraphics[width=\textwidth]{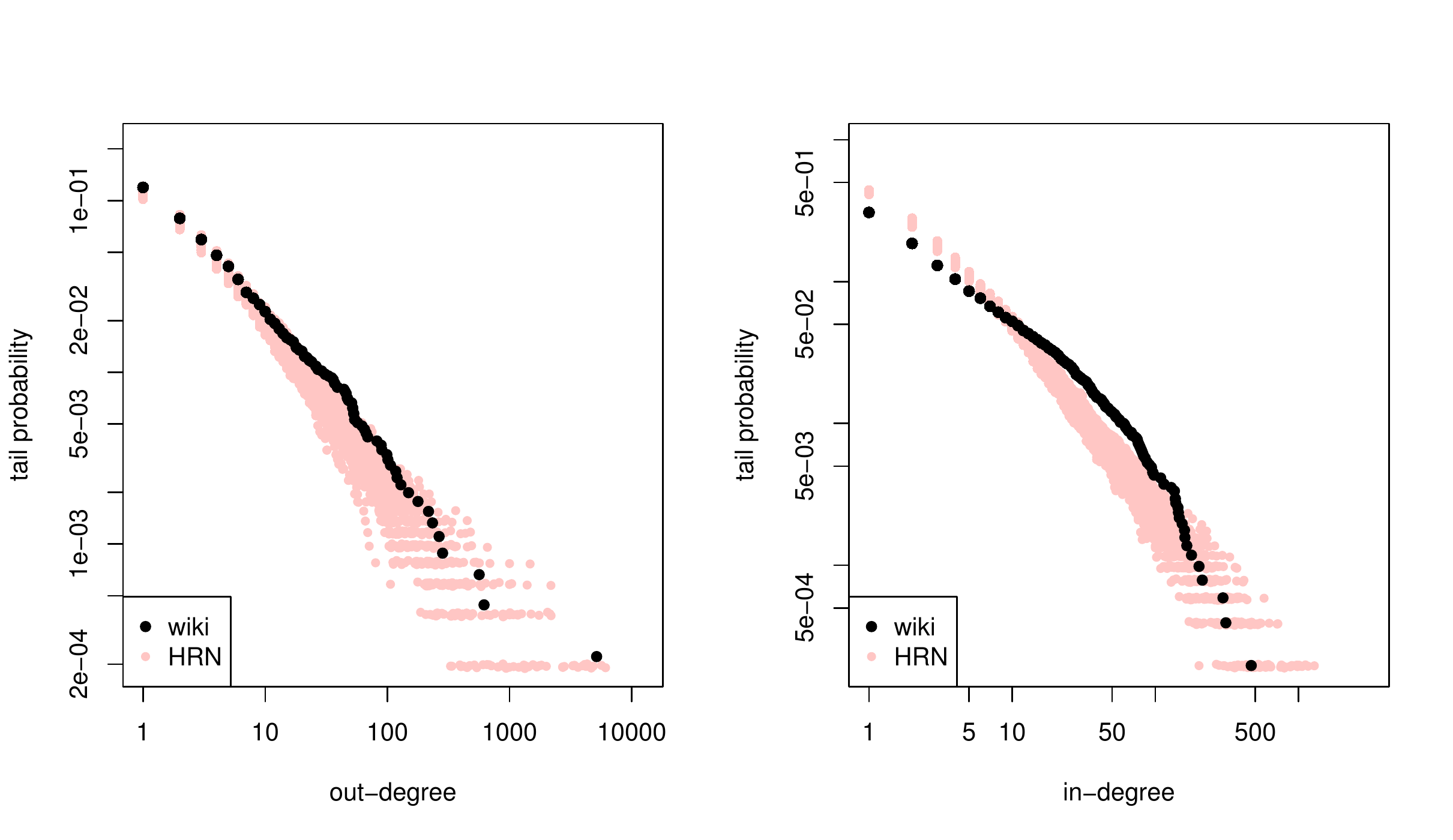}
	\caption{Out-degree and in-degree tail distributions of 50 
		independent HRNs (red), compared with those from the 
		Wikipedia data (black).}
	\label{fig:overlay_wiki}
\end{figure}

In addition, we generate $50$ independent replications of the HRN 
using the 
given estimates. The out-degree and in-degree tail distributions of 
these simulated networks are collected in 
Figure~\ref{fig:overlay_wiki}. 
The left panel shows that the out-degree tail distribution of the 
selected sub-network is well covered by the overlaid plot of the 
simulated counterparts, whereas the right reveals small but 
acceptable discrepancy between the overlaid plot and the in-degree 
distribution of the real data. Overall, our analysis suggests 
the evolution of the selected sub-network of the Dutch 
Wikipedia talk data follows a linear PA mechanism, 
which also confirms the flexibility of the proposed hybrid model.

\subsection{Facebook Wall Posts}
\label{Sec:fb}
The Facebook wall post dataset collects data from a regional network 
of
users in New Orleans from 09/13/2004 to 01/21/2009. The data forms a
directed graph where the nodes are Facebook users and each directed 
edge represents a post from one node to another node's page. 
Like the previous one, this dataset contains three columns, where 
the first two contain the identifiers of the individual users, while 
the third records the timestamp of the corresponding post. 

\begin{figure}[tbp]
	\centering
	\includegraphics[width=\textwidth]{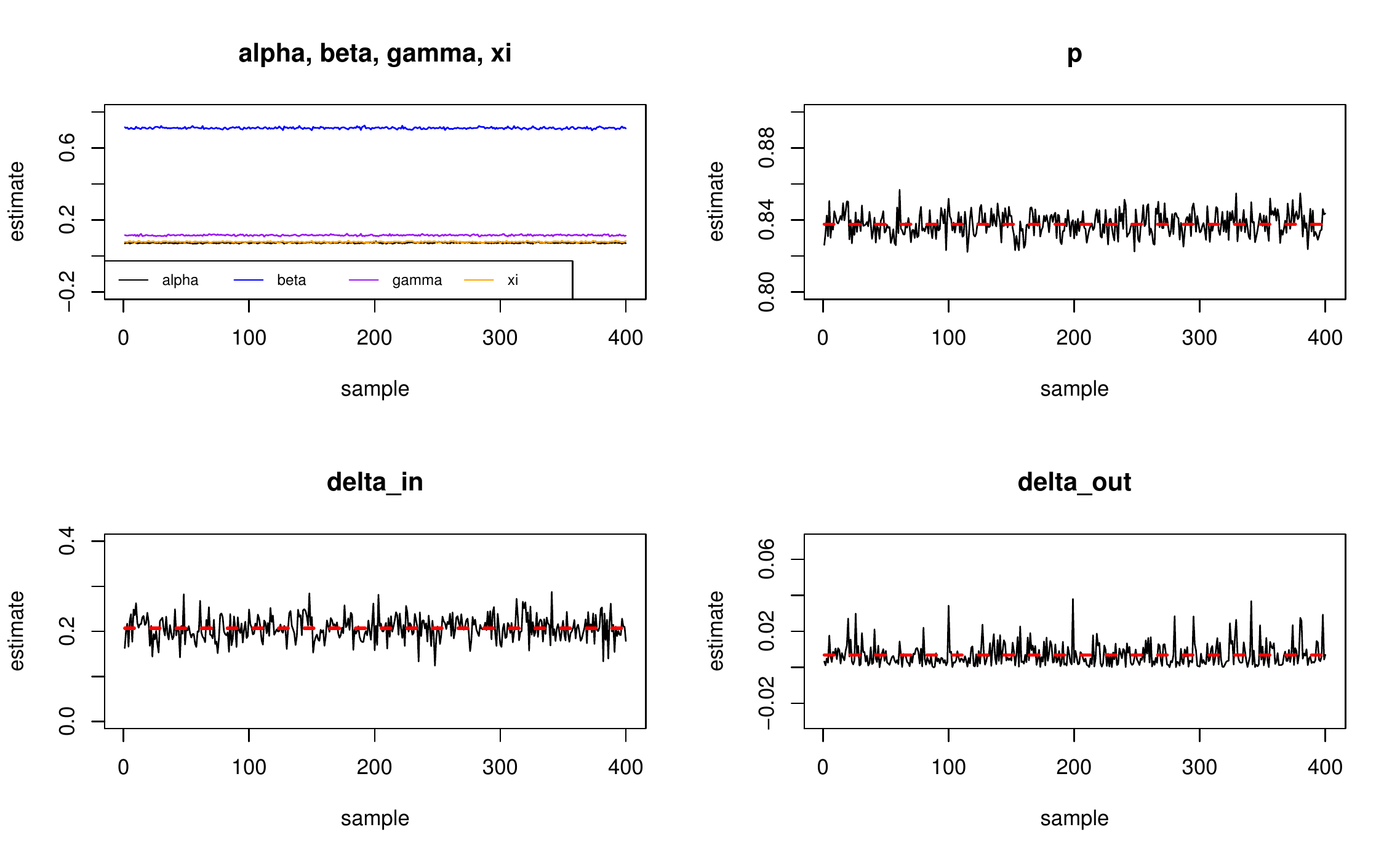}
	\caption{Graphical diagnostics of the convergence for the M-H 
		algorithm applied to the sub-network of Facebook wall posts 
		dataset; red dashed lines represent the estimated values.}
	\label{fig:MH_fb}
\end{figure}

We select the sub-network based on the 
timestamp from 01/01/2006 to 06/31/2006, which
consists of 4,200 nodes and 11,422 directed edges. Fitting the 
proposed hybrid model to the data, we get
\[\widehat{\boldsymbol{\theta}}_{\rm FB}:=(\hat{\alpha}, 
\hat{\beta}, \hat{\gamma}, \hat{\xi}, \hat{p}, 
\hat{\delta}_{\rm in}, \hat{\delta}_{\rm out}) = (0.071, 0.714, 
0.114, 0.077, 0.830, 0.172, 0.007).\]
The estimates are obtained via the integration of the N-M and M-H 
algorithms, where the settings of burn-in number, iteration size and 
thinning gap are identical to the previous example. In 
Figure~\ref{fig:MH_fb}, we verify the convergence of the estimates 
based on the M-H algorithm. Once again, we do not observe 
significant difference between the corresponding estimates from the 
two algorithms. 

\begin{figure}[tbp]
	\centering
	\includegraphics[width=\textwidth]{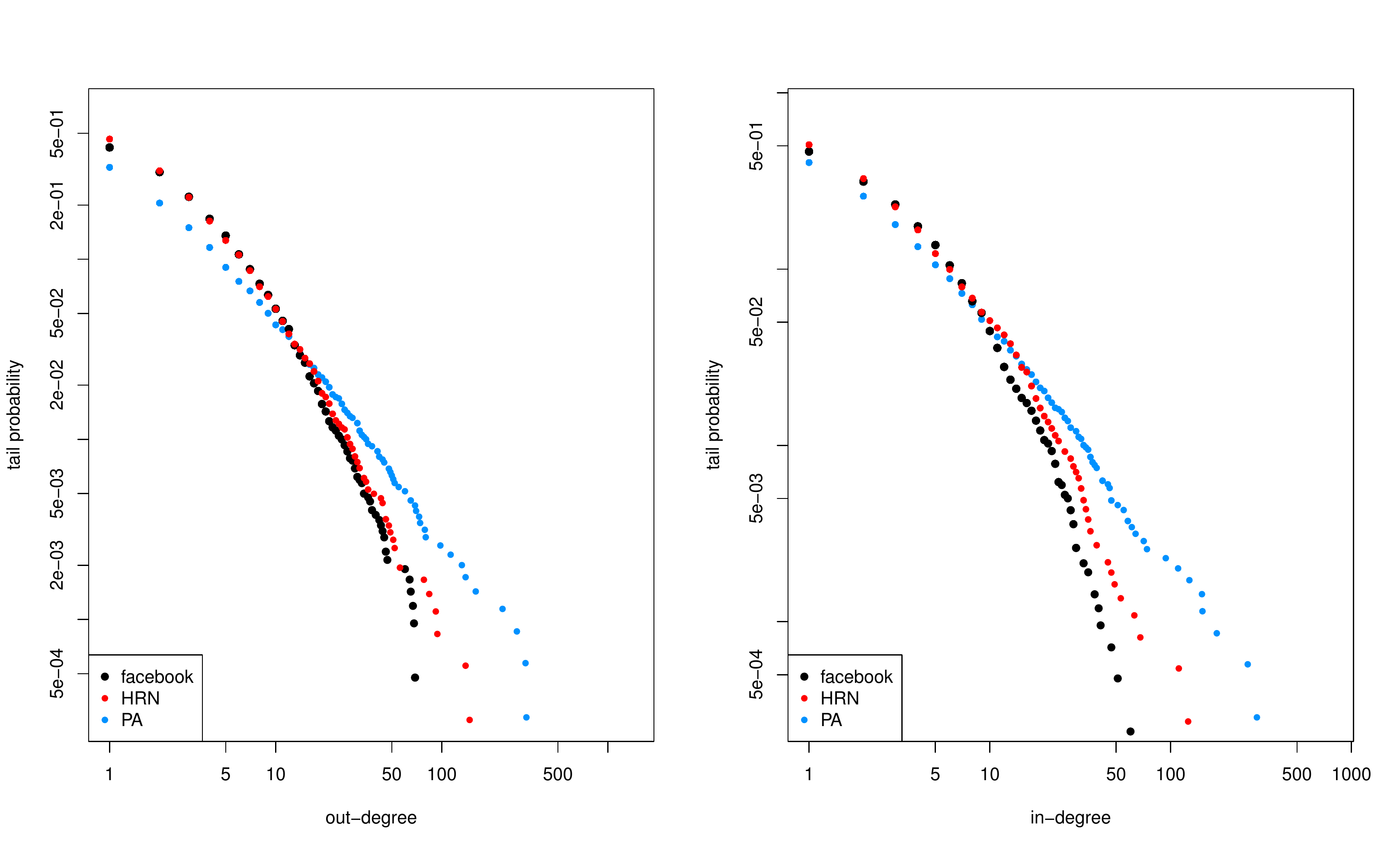}
	\caption{Out-degree and in-degree tail distributions of HRNs 
		(red), 
		directed PA networks (blue) and the sub-network of Facebook 
		wall posts (black).}
	\label{fig:tail_fb}
\end{figure}

We assess the goodness-of-fit of the model through the tail 
distributions of in-degree and out-degree. Specifically, we 
generate an HRN using $\widehat{\boldsymbol{\theta}}_{\rm FB}$, and 
compare the 
empirical out- and in-degree tail distributions from the simulated 
network 
with those from the real data. Results are presented in 
Figure~\ref{fig:tail_fb}. For comparison, we also fit a pure PA 
model to
the same sub-network data using the estimation method derived
in~\citet{Wan2017}, and plot the corresponding empirical tail 
distributions (blue dots) 
in Figure~\ref{fig:tail_fb}.

The empirical tail distributions show that the HRN provides a better 
fit to the network data by introducing the extra UA rules controlled 
by the paramter $p$. 
Empirical out- and in-degree distributions from the pure PA model 
display 
heavier tails than those from the Facebook dataset, leading to
a clear deviation from the data. 
In the HRN, however, the existence of $100(1-p)\%$ edges created by 
the UA rule has reduced the heaviness of the empirical tails, thus 
reducing the discrepancy between the empirical distributions.

\begin{figure}[tbp]
	\centering
	\includegraphics[width=\textwidth]{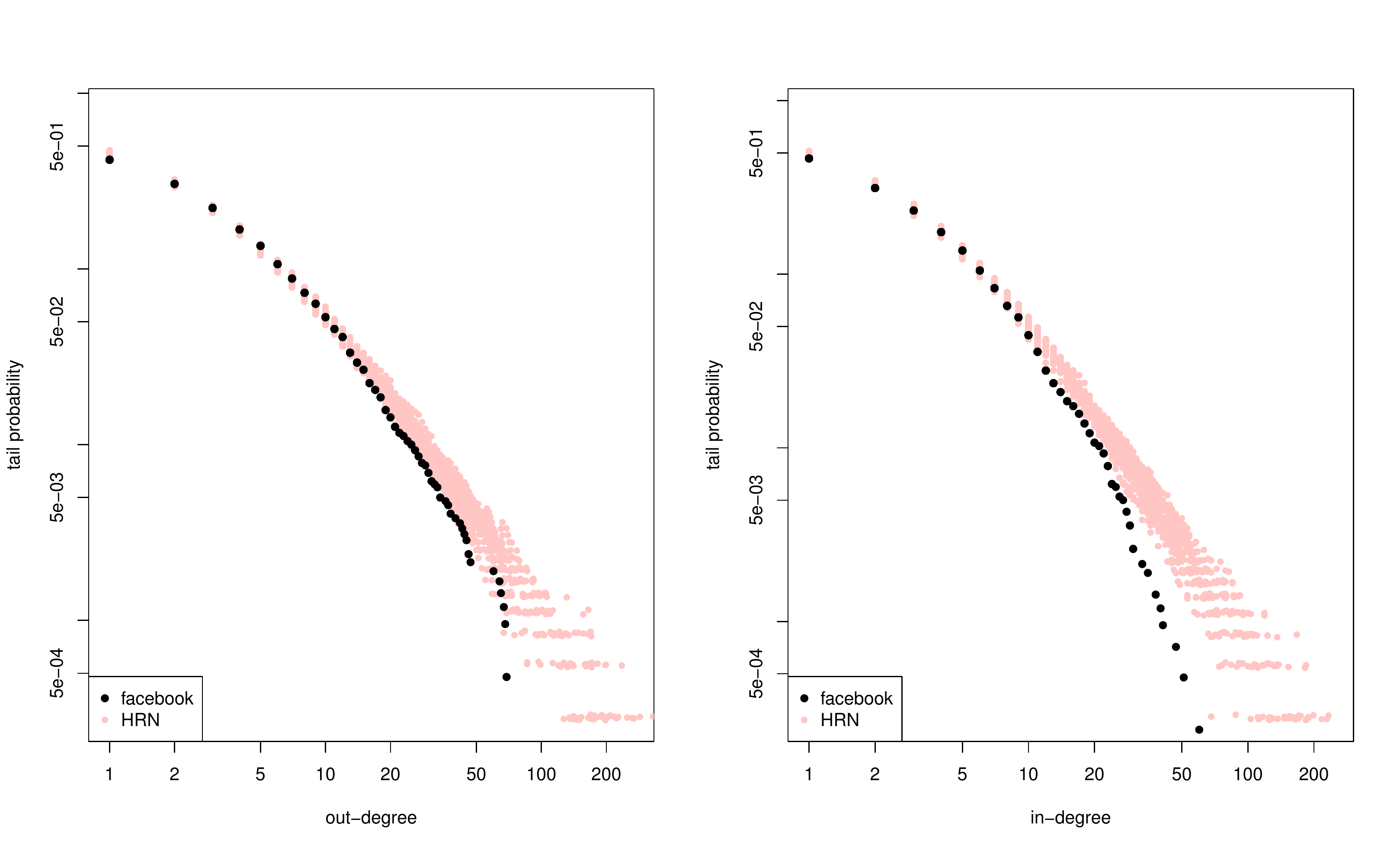}
	\caption{Out-degree and in-degree tail distributions of 50 
		independent HRNs (red), compared with those from the 
		Facebook data (black).}
	\label{fig:overlay_fb}
\end{figure}

Furthermore, we generate $50$ independent HRNs by using 
$\widehat{\boldsymbol{\theta}}_{\rm FB}$, and overlay their 
empirical out-degree and in-degree 
distributions in Figure~\ref{fig:overlay_fb}. The graphical results 
show that 
the out-degree tail distribution is better captured by the HRN than 
the 
in-degree distribution, as it appears on the lower bound of the 
overlaid tail distributions of the simulated networks. The 
in-degree tail distribution of the Facebook sub-network is not 
covered by the counterparts of the simulated networks, though the 
shapes look similar and the deviation is not large.

The discrepancy in Figure~\ref{fig:overlay_fb} may be due to 
the reciprocity feature in the Facebook wall posts.
The wall posts activities among the Facebook users in a specific 
region 
tend to be reciprocated: when a friend posts a message on one's wall,
he/she is likely to reply quickly. 
In fact, using the \verb|reciprocity()| function in the 
\verb|igraph| package~\citep{Csardi2006},
we see that the proportion of reciprocated edges in the sub-network 
is
over $0.18$. 
Indeed, the reciprocated wall posts
are certainly not uniform, thus not very well characterized by the 
parameter $p$.
To better study the reciprocity feature, we may consider other
variants of the PA model, which are left as future work.

\section{Discussion}
\label{Sec:discus}

In this paper, we propose a class of hybrid model simultaneously 
presenting the preferential attachment (PA) and uniform attachment 
(UA) mechanisms, which are governed by a tuning parameter $p$. 
Two standard methods, the Nelder-Mead (N-M) algorithm and the 
Metropolis-Hastings (M-H) algorithm, are adopted for parameter 
estimation. Through extensive simulations and a sensitivity study, 
we find that the N-M algorithm is preferred, but the corresponding 
success rate of producing estimation result depends heavily on 
the selection of the initial simplex. We thus consider an integrated 
approach where we use the more robust M-H algorithm to get the 
initial values for the target parameters, followed by the 
implementation of the N-M algorithm. 

In addition, we fit the HRN model to two real network datasets: the 
Dutch 
Wikipedia talk and Facebook wall posts, where we see that the 
proposed hybrid model provides a more flexible modeling framework 
compared with the directed PA network model as in~\citep{Wan2017}. 
The extra tuning parameter $p$ helps correct the 
tail distributions of out- and in-degrees. 

From the Facebook example, it is worth noting that even though the 
heaviness 
of the tail distributions (for both out- and in-degrees) has 
been weakened by $p$ in the HRN, the proposed UA part is not 
able to fully capture the reciprocity property in the real network. 
We here provide three classes of possible remedies: (1) It may be 
worthwhile to 
try some algorithmic approaches, such as network rewiring, to 
wash off the nodes of large in-degree or out-degree in the 
simulated networks; (2) We may consider modifying the model directly 
by 
introducing another parameter measuring the rate of reciprocation; 
(3)
We may consider a more realistic mixer (some suitable light-tailed 
distribution) rather than simple UA in the present hybrid setting. 
We will report our research outcomes elsewhere in the future.

\appendix
\section{Proof of Theorem~\ref{thm:focus}}
\label{sec:appthm1}

We explicitly demonstrate the derivations of the focusing theorem 
for the in-degree distribution, and the methodology is also 
applicable to 
out-degrees. 
Let $\mathcal{F}_n$ denote the $\sigma$-field generated by the 
network evolution up to 
$n$ steps. Set $\tau$ to be an $(\mathcal{F}_n)_{n\ge 0}$-stopping 
time, then 
\[
\mathcal{F}_{\tau} = \{F: F\cap\{\tau=n\}\in \mathcal{F}_n\}.
\]
For $i\ge 1$, let $S_i$ be the time when node $i$ is created, i.e.,
\[
S_i = \inf\{n\ge 0: |V_n| = i\}.
\]
Then $S_i$ is an $(\mathcal{F}_n)_{n\ge 0}$-stopping time.
Also, for $n\ge k\ge 0$, we have
\[
\{S_i+k=n\} = \{S_i=n-k\}\in \mathcal{F}_{n-k}\subset \mathcal{F}_n,
\]
so $S_i+k$, $k\ge 0$, is a stopping time with respect to 
$(\mathcal{F}_n)_{n\ge 0}$.

Since the in-degree of $i$ is increased at 
most by 
$1$ at each evolutionary step, the probability of 
the event $\{\Din_i(S_i+n + 1) = \Din_i(S_i+n)\}$ under
$\Prob^{\field_{S_i+n}}(\cdot):=\Prob(\cdot|\field_{S_i+n})$ is
\begin{align*}
	\Prob^{\field_{S_i+n}}&\left(\Din_i(S_i+n + 1) = 
	\Din_i(S_i+n)\right) \\
	&= 1 - (\alpha + \beta) \left[\frac{p(\Din_i(S_i+n) + 
		\deltain)}{\sum_{k \in V_{S_i+n}} (\Din_k(S_i+n) + 
		\deltain)} + 
	\frac{(1 - p)}{|V_{S_i+n}|}\right].
\end{align*}
Therefore,
\begin{align}
	&\Expe^{\field_{S_i+n}}\left[\Din_i(S_i+n + 1) \right] 
	\nonumber \\
	&{}= \Din_{i}(S_i+n) + (\alpha + \beta) 
	\left[\frac{p(\Din_i(S_i+n) + 
		\deltain)}{\sum_{k \in V_{S_i+n}} (\Din_k(S_i+n) + 
		\deltain)} + 
	\frac{(1 - p)}{|V_{S_i+n}|}\right]\nonumber\\
	& = \Din_{i}(S_i+n) + (\alpha + \beta) 
	\left[\frac{p(\Din_i(S_i+n) + 
		\deltain)}{S_i+n+1+\deltain |V_{S_i+n}|} + 
	\frac{(1 - p)}{|V_{S_i+n}|}\right].
	\label{eq:indegcondexp}
\end{align}

Write
$C_1 := p(\alpha + \beta)/(1 + \deltain(1 - \beta))$, 
and $\tdeltain := \deltain / p + (1 - p)/(p(1 - 
\beta))$, then we have
\begingroup
\allowdisplaybreaks
\begin{align}
	&\Expe\left[\Din_i(S_i+n + 1) + \tdeltain\right] \nonumber
	\\&\quad{}= 
	\Expe\left[\Din_{i}(S_i+n) + \tdeltain + (\alpha + \beta) 
	\left(\frac{p(\Din_i(S_i+n) + \deltain)}{S_i + n+1 + 
		\deltain|V_{S_i+n}|} 
	+ 
	\frac{(1 - p)}{|V_{S_i+n}|}\right)\right]\nonumber
	\\ &\quad{}= \Expe\left[\Din_{i}(S_i+n) + \tdeltain + (\alpha + 
	\beta) 
	\left(\frac{p(\Din_i(S_i+n) + \tdeltain)}{S_i + n + 1+
		\deltain|V_{S_i+n}|} 
	+ \frac{(1 - p)}{|V_{S_i+n}|} \right.\right.\nonumber
	\\ &\qquad\qquad{} - \left.\left.  \frac{p(\tdeltain - 
		\deltain)}{S_i+n + 1 + \deltain|V_{S_i+n}|} 
	\right)\right]\nonumber
	\\ &\quad{}= \Expe\left[(\Din_{i}(S_i+n) + \tdeltain)\left(1 + 
	\frac{p(\alpha +\beta)}{S_i + n+1 + 
		\deltain|V_{S_i+n}|}\right)\right] \nonumber
	\\ &\qquad{}\qquad{}+ \Expe\left[\frac{1 - p}{|V_{S_i+n}|} - 
	\frac{p(\tdeltain - \deltain)}{S_i+n + 1 + 
		\deltain|V_{S_i+n}|}\right]\nonumber
	\\ &\quad{}\le \Expe\left[(\Din_{i}(S_i+n) + \tdeltain) \left(1 
	+ 
	\frac{p(\alpha +\beta)}{n+1 + \deltain|V_{n}|}\right)\right] 
	\nonumber\\
	&\qquad\qquad{}+ \frac{1 - p}{n(1 - \beta)} 
	\Expe\left|(1 - \beta)\frac{S_i+n + 1}{|V_{S_i+n}|} - 1
	\right|\nonumber\\
	&\quad= \Expe\left[(\Din_{i}(S_i+n) + \tdeltain) \left(1 + 
	\frac{C_1}{n}\right)\right]\nonumber \\
	&\qquad\qquad{}+ \Expe\left[(\Din_{i}(S_i+n) + \tdeltain) \left(
	\frac{p(\alpha +\beta)}{n+1 + 
		\deltain|V_{n}|}-\frac{C_1}{n}\right)\right]
	\nonumber\\
	&\qquad\qquad\qquad{}+ \frac{1 - p}{n(1 - \beta)} \Expe\left|(1 
	- 
	\beta)\frac{S_i+n + 1}{|V_{S_i+n}|} - 1
	\right|
	\nonumber\\
	&\quad=: I_i(n)+II_i(n)+III_i(n).\label{eq:Din_upper}
\end{align}
\endgroup
For $II_i(n)$, we have
\begin{align*}
	II_i(n)&\le \Expe\left[(\Din_{i}(S_i+n) + \tdeltain) \left|
	\frac{p(\alpha +\beta)}{n+1 + 
		\deltain|V_{n}|}-\frac{C_1}{n}\right|\right]\\
	& \le \Expe\left[(\Din_{i}(S_i+n) + \tdeltain)
	\frac{\deltain 
		\bigl||V_n|-(1-\beta)n\bigr|+1}{(1+\deltain(1-\beta))n^2}\right].
\end{align*}
Since for $n\ge 1$, $|V_n|-1$ is a binomial random variable with 
success probability $1-\beta$,
we apply the Chernoff bound to get
\begin{equation}\label{eq:chernoff}
	\Prob\left(\left||V_n|  -(1-\beta)n\right| \ge \sqrt{12(1 - 
		\beta) n \log{n}}\right) \le \frac{2}{n^4},
\end{equation}
and noticing that $\left||V_n|  -(1-\beta)n\right|\le n$, and 
$\Din_i(S_i+n)\le n+1$, we have
\begin{align*}
	II_i(n)&\le \left(\frac{\deltain\sqrt{12n\log 
			n}}{n^2}+\frac{1}{n^2}\right)\Expe(\Din_{i}(S_i+n) + 
			\tdeltain)\\
	&\qquad+ \frac{2(\deltain n+1) (n+1+\tdeltain)}{n^6}\\
	&\le \left(\frac{\deltain\sqrt{12n\log 
			n}}{n^2}+\frac{1}{n^2}\right)\Expe(\Din_{i}(S_i+n) + 
			\tdeltain)+ 
	\frac{2(1+\deltain) (2+\tdeltain)}{n^4}.
\end{align*}
For $III_i(n)$, we apply the Cauchy-Schwartz inequality to obtain
\begin{align*}
	III_i(n)&\le \frac{1-p}{n(1-\beta)} \left(\Expe 
	\left[\left((1-\beta)(S_i+n+1)-|V_{S_i+n}|\right)^2\right]\right)^{1/2}\left(\Expe\left[|V_{n}|^{-2}\right]\right)^{1/2}.
\end{align*}
By Theorem 3.9.4 in \cite{athreya:ney:2004}, we have as $n\to\infty$,
\[
n^2\Expe\left[|V_{n}|^{-2}\right]\to (1-\beta)^{-2}.
\]
Meanwhile, 
\begin{align*}
	\Expe 
	&\left[\left((1-\beta)(S_i+n+1)-|V_{S_i+n}|\right)^2\right]\\
	&= \text{Var}\left(|V_{S_i+n}|\right)+ 
	\text{Var}\left((1-\beta)(S_i+n+1)\right)\\
	&= \beta(1-\beta)\left(\Expe\left(S_i\right)+n\right) + 
	2(1-\beta)^2\text{Var}(S_i).
\end{align*}
Hence, there exists some constant $A_i>0$ such that
\[
III_i(n)\le A_i n^{-3/2}.
\]
Then it follows from \eqref{eq:Din_upper} that
\begin{align*}
	\Expe&\left[\Din_i(S_i+n + 1) + \tdeltain\right] \\
	&\le \Expe\left[(\Din_{i}(S_i+n) + \tdeltain) \left(1 + 
	\frac{C_1}{n}+\frac{1+\deltain\sqrt{12n\log 
			n}}{n^2}\right)\right]\\
	&\quad + 2(1+\deltain)(2+\tdeltain)n^{-4}+A_i n^{-3/2}.
\end{align*}

Iterating backwards for $n$ times gives
\begingroup
\allowdisplaybreaks
\begin{align*}
	&\Expe\bigl[\Din_i(S_i+n + 1) + \deltain\bigr] 
	\\ &\quad{}\le 
	\bigl(\deltain \alpha + (1 + \deltain) \gamma\bigr) \prod_{r 
		=1}^{n} \left(1 + 
	\frac{C_1}{r} + \frac{1+\sqrt{12(1 - \beta)r \log r}}{r^2}\right)
	\\&\qquad{}+ \sum_{r = 1}^{n}\frac{2(1 + \deltain)(2 +
		\tdeltain)}{r^2} \prod_{s = r + 1}^{n} \left(1 + 
	\frac{C_1}{s} +
	\frac{1+\sqrt{12(1 - \beta)s \log s}}{s^2}\right).
\end{align*}
\endgroup
Note that 
\begin{align*}
	&\prod_{s = r + 1}^{n} \left(1 + \frac{C_1}{s} +
	\frac{1+\sqrt{12(1 - \beta)s \log s}}{(1 + \deltain(1 - 
		\beta))s^2}\right) 
	\\&\quad{}\le \exp\left\{\sum_{s = r + 1}^{n} 
	\left(\frac{C_1}{s} +
	\frac{1+\sqrt{12(1 - \beta)s \log s}}{(1 + \deltain(1 - 
		\beta))s^2}\right) \right\}\\
	&\quad{}\le (n/r)^{C_1}\exp\left\{\sum_{s = 1}^{\infty} 
	\frac{1+\sqrt{12(1 - \beta)s \log s}}{(1 + \deltain(1 - 
		\beta))s^2}\right\},
\end{align*}
which gives
$$\sup_{i \ge 1} \frac{\Expe[\Din_i(n)]}{n^{C_1}}\le \sup_{i \ge 1} 
\frac{\Expe[\Din_i(S_i+n)+\tdeltain]}{n^{C_1}} < \infty.$$
This completes the proof.

\section{Proof of Theorem~\ref{thm:mtg}}
\setcounter{equation}{0}
\label{sec:appthm2}

Like the proof of Theorem~\ref{thm:focus}, we only present the 
details for the 
in-degree sequence. The derivations for out-degree sequence are 
similar, so omitted.

For $k, \ell \ge 0$, we have
\begin{align*}
	\Expe\left[M^{\rm in}_{k + \ell} - M^{\rm in}_k\right] &= 
	\sum_{s = k + 1}^{k + \ell} 
	\Expe\left[M^{\rm in}_{s} - M^{\rm in}_{s - 1}\right]
	\\ &= \sum_{s = k + 1}^{k + \ell} 
	\Expe\left[\frac{(1 - p)(\alpha + 
		\beta)/|V_{S_i+s}|}{\prod_{\ell 
			= 1}^{s} \left(1 + \frac{p(\alpha + \beta)}{S_i+\ell + 1 
			+ 
			\deltain|V_{S_i+\ell}|}\right)}\right]
	\\&\le \sum_{s = k + 1}^{k + \ell} \Expe\left[ \prod_{\ell = 
		1}^{s} \left(1 + \frac{p(\alpha + \beta)}{(1 + 
		\deltain)(S_i+\ell + 1)}\right)\frac{(1 - p)(\alpha + 
		\beta)}{|V_{S_i+s}|}\right]\\
	&\le \sum_{s = k + 1}^{k + \ell}  \prod_{\ell = 
		1}^{s} \left(1 + \frac{p(\alpha + \beta)}{(1 + 
		\deltain)(\ell + 1)}\right)\Expe\left[\frac{(1 - p)(\alpha + 
		\beta)}{|V_{s}|}\right],
\end{align*}
where 
$$\prod_{\ell = 
	1}^{s} \left(1 + \frac{p(\alpha + \beta)}{(1 + 
	\deltain)(\ell + 1)}\right) = \frac{\Gamma\left(s + 2 + 
	\frac{p(\alpha + \beta)}{1 + \deltain}\right)}{\Gamma(s + 
	2)\Gamma\left(2 + \frac{p(\alpha + \beta)}{1 + 
		\deltain}\right)} \sim s^{-(p(\alpha +\beta))/(1 + 
	\deltain)},$$
for $s$ large. For the expectation in the summand, we have
$$\Expe\left[\frac{(1 - p)(\alpha +	\beta)}{|V(s)|}\right] = (1 
- p)(\alpha + \beta) \left(\frac{1}{(1 - \beta)s} + 
\Expe\left[\frac{1}{|V(s)|} - \frac{1}{(1 - 
	\beta)s}\right]\right).$$
By the Chernoff bound in \eqref{eq:chernoff}, we get
\begin{align*}
	\Expe\left[\left|\frac{1}{|V(s)|} - \frac{1}{(1 - 
		\beta)s}\right|\right] &= \Expe\left[\frac{\bigl||V(s)| - (1 
		- 
		\beta)s\bigr|}{(1 - \beta)s |V(s)|}\right]
	\\ &\le \frac{\sqrt{12(1 - \beta) s \log s}}{(1 - \beta) s |(1 - 
		\beta)s - \sqrt{12(1 - \beta)s\log{s}}|}
	\\&\qquad{}+ \frac{(2 - \beta) s}{(1 - \beta)s\sqrt{12(1 - 
			\beta)s \log s}} \times \frac{2}{s^4}.
\end{align*}
Putting them together, we conclude that $\Expe\left|M^{\rm in}_{k + 
	\ell} - M^{\rm in}_k\right| 
\to 0$ as $k \to \infty$, suggesting that $\left\{\Expe\left[M^{\rm 
	in}_n\right]\right\}_{n \ge 1}$ is a \emph{cauchy} sequence in 
$L_1$ space. Applying the {martingale convergence 
	theorem}~\citep[Theorem~4.2.11]{Durrett2006} gives that there 
exists some 
finite random variable $L_i$ such that as $n\to\infty$,
\begin{equation}\label{eq:min_conv}
	M^{\rm in}_n \stackrel{a.s.}{\longrightarrow} L_i.
\end{equation}

Then it remains to show the almost sure convergence of 
\[
X_n := 
\prod_{k=0}^{n-1}\left(1+\frac{(\alpha+\beta)p}{S_i+k+1+\deltain|V_{S_i+k}|}\right).
\]
Consider $\log X_n$, and rewrite it as
\begin{align}
	\log X_n =& 
	\left[\sum_{k=0}^{n-1}\log\left(1+\frac{(\alpha+\beta)p}{S_i+k+1+\deltain|V_{S_i+k}|}\right)
	-\sum_{k=0}^{n-1}\frac{(\alpha+\beta)p}{S_i+k+1+\deltain|V_{S_i+k}|}\right]\nonumber\\
	&+ 
	\left[\sum_{k=0}^{n-1}\frac{(\alpha+\beta)p}{S_i+k+1+\deltain|V_{S_i+k}|}
	-\sum_{k=1}^{n}\frac{C_1}{S_i+k+1}\right]\nonumber\\
	&+\left[
	\sum_{k=1}^{n}\frac{C_1}{S_i+k}
	- C_1\log n
	\right]\nonumber\\
	&+ C_1\log n =: P_1(n)+P_2(n)+P_3(n)+C_1\log n.\label{eq:Xn}
\end{align}
For $P_1(n)$, we first note that $\log(1+x)-x\le 0$ for all $x\ge 
0$. Then 
$P_1(n+1)-P_1(n)\le 0$, i.e. $P_1(n)$ is decreasing in $n$, and it 
suffices to show
\[
P_1(\infty) := 
\sum_{k=0}^{\infty}\left(\log\left(1+\frac{(\alpha+\beta)p}{S_i+k+1+\deltain|V_{S_i+k}|}\right)
-\frac{(\alpha+\beta)p}{S_i+k+1+\deltain|V_{S_i+k}|}\right)
\]
is finite almost surely. Note also that $|\log(1+x)-x|\le x^2/2$, 
for all $x\ge 0$, then 
\begin{align*}
	\Expe&\left|\sum_{k=0}^{\infty}\left(\log\left(1+\frac{(\alpha+\beta)p}{S_i+k+1+\deltain|V_{S_i+k}|}\right)
	-\frac{(\alpha+\beta)p}{S_i+k+1+\deltain|V_{S_i+k}|}\right)\right|\\
	&\le 
	\sum_{k=0}^{\infty}\Expe\left|\log\left(1+\frac{(\alpha+\beta)p}{S_i+k+1+\deltain|V_{S_i+k}|}\right)
	-\frac{(\alpha+\beta)p}{S_i+k+1+\deltain|V_{S_i+k}|}\right|\\
	&\le \frac{(\alpha+\beta)^2 
		p^2}{2}\sum_{k=0}^{\infty}\Expe\left(\frac{1}{S_i+k+1+\deltain|V_{S_i+k}|}\right)^2\le
	\sum_{k=1}^{\infty}\frac{1}{k^2}<\infty.
\end{align*}
Hence, $P_1(n)\convas P_1(\infty)$.

For $P_2(n)$, we apply \cite[Theorem 3.9.4]{athreya:ney:2004} to 
conclude that there exists a finite r.v. $Z$ such that
\[
\sum_{k=1}^\infty\left(\frac{(\alpha+\beta)p}{k+\deltain 
	|V(k-1)|}-\frac{C_1}{k}\right)\convas Z,
\]
then 
$$P_2(n)\convas 
Z-\sum_{k=1}^{S_i}\left(\frac{(\alpha+\beta)p}{k+\deltain 
	|V(k-1)|}-\frac{C_1}{k}\right)
=: P_2(\infty).$$
Meanwhile,
since $\sum_{k=1}^n 1/k -\log n\to \tilde{c}$ as $n\to\infty$, where 
$\tilde{c}$ is Euler's constant,
then for $i\ge 1$,
\[
P_3(n)\convas 
C_1\left(\tilde{c}+\log(S_i+1)-\sum_{k=1}^{S_i}\frac{1}{k}\right)=: 
P_3(\infty).
\]
Then we conclude from \eqref{eq:Xn} that
\begin{equation}\label{eq:Xn_limit}
	\frac{X_n}{n^{C_1}} \convas 
	\exp\left(P_1(\infty)+P_2(\infty)+P_3(\infty)\right) <\infty.
\end{equation}
The results in the theorem follow by combining \eqref{eq:Xn_limit} 
with \eqref{eq:min_conv}, where we set 
\[
\zeta_i := L_i 
\exp\left(-(P_1(\infty)+P_2(\infty)+P_3(\infty))\right).
\]

\section{Proof of Theorem~\ref{thm:degnum}}
\setcounter{equation}{0}
\label{sec:appthm3}

Analogous to the previous proofs, we present the major steps of the 
proof for in-degree. 
Applying the argument as in the proof of \cite[Proposition 
8.4]{vanderhofstad2017},
we have the concentration result that
$$
\frac{N^\text{in})_m(n)}{n}-\frac{\Expe\left(N^\text{in}_m(n)\right)}{n}\stackrel{p}{\longrightarrow}0.
$$ 
Then it suffices to find the asymptotic limit of 
$\Expe\left(N^\text{in}_m(n)\right)/{n}$.

We consider the approximation of the attachment 
probability:
\[\frac{p \bigl(\Din_{i}(n) + \deltain \bigr)}{\bigl(1 + \deltain(1 
	- \beta) \bigr)n} + \frac{1 - p}{(1 - \beta)n} = 
\frac{\Din_{i}(n) + 
	\deltain + \frac{1 - p}{p(1 - \beta)}\bigl(1 + \deltain(1 - 
	\beta)\bigr)}{\bigl(1 + \deltain (1 - \beta)\bigr) n/p}.\]
Recall that
\[\tilde{\delta}_{\rm in} = \deltain + \frac{1 - p}{p(1 - 
	\beta)}\bigl(1 + \deltain(1 - 
\beta)\bigr) = \frac{\deltain}{p} + \frac{1 - p}{p(1 - \beta)}.\] 
Using Chernoff bound in \eqref{eq:chernoff1}, we have
\begin{align}
	&\left|\Expe\left[\frac{p\bigl(\Din_i(n) + \deltain \bigr)}{n + 
		1 
		+ |V_n|\deltain} + \frac{1 - p}{|V_n|}\right] - 
	\Expe\left[\frac{\Din_{i}(n) + \tilde{\delta}_{\rm in}}{\bigl(1 
		+ \deltain (1 - \beta)\bigr) n/p}\right]\right| \nonumber 
	\\ &\quad{}\le C n^{-3/2}\sqrt{\log{n}}, \label{eq:chernoff1}
\end{align}
for some constant $C > 0$. Consider a in-degree sequence 
$\left\{\tilde{D}_i^{\rm in}(n)\right\}$ from a 
directed PA network with set of parameters $(\alpha, \beta, \gamma, 
\tilde{\delta}_{\rm in}, \tilde{\delta}_{\rm out})$, as studied in
~\citet{resnick:samorodnitsky:towsley:davis:willis:wan:2016, 
	Wan2017}. Establish an argument similar to 
Equation~\eqref{eq:chernoff} as follows:
\[\left|\Expe\left[\frac{p\bigl(\Din_i(n) + \deltain \bigr)}{n + 
	1 + |V_n|\deltain} + \frac{1 - p}{|V_n|}\right] - 
\Expe\left[\frac{\tilde{D}_i^{\rm in}(n) - \tilde{\delta}_{\rm 
		in}}{n + 1 + |V_n|\tilde{\delta}_{\rm in}}\right]\right| 
\nonumber 
\le \tilde{C} n^{-3/2}\sqrt{\log{n}},
\]
for some constant $\tilde{C} > 0$.
Note
\[\Prob\bigl(\Din_i(n) = m\bigr) = \sum_{j = m - 1}^{m} 
\Prob\bigl(\Din_i(n) = m \given \Din_i(n - 1) = 
j\bigr)\Prob\bigl(\Din_i(n - 1) = j\bigr).\]
By the developed Chernoff bounds, we have
\[\Prob\bigl(\Din_i(n) = m\bigr) \le \Prob\bigl(\tilde{D}_i^{\rm 
	in}(n) = m\bigr) + (C + \tilde{C}) \sum_{k = i}^{n} k^{-3/2} 
\sqrt{\log{k}}.\]
Noticing that $\sum_{k = i}^{n} k^{-3/2} 
\sqrt{\log{k}} < \infty$ as $n \to \infty$, we complete the proof by 
applying the results derived in~\citet{Wang2020}.

\end{document}